\documentclass{sig-alternate}
\usepackage{graphicx}
\usepackage{times}
\usepackage{algorithm}
\usepackage{epsfig}
\usepackage{subfigure}
\usepackage{algorithmic}

\usepackage[hyphens]{url}
\usepackage[hyphenbreaks]{breakurl}

\usepackage{color}

\newtheorem{theorem}{Theorem}
\newcommand{\topic}[1]{\vspace{5pt} \noindent \underline{\bf #1}}
\newcommand{\emtopic}[1]{\vspace{3pt} \noindent \underline{\em #1}}

\newcommand{\commentout}[1]{}
\newcommand{\eat}[1]{}
\newcommand{\eatamol}[1]{}

\newcommand{\squishlist}{
 \begin{list}{$\bullet$}
  { \setlength{\itemsep}{0pt}
     \setlength{\parsep}{3pt}
     \setlength{\topsep}{3pt}
     \setlength{\partopsep}{0pt}
     \setlength{\leftmargin}{0.5em}
     \setlength{\labelwidth}{1em}
     \setlength{\labelsep}{0.5em} } }

\newcommand{\squishlisttwo}{
 \begin{list}{$\bullet$}
  { \setlength{\itemsep}{0pt}
     \setlength{\parsep}{0pt}
    \setlength{\topsep}{0pt}
    \setlength{\partopsep}{0pt}
    \setlength{\leftmargin}{0.5em}
    \setlength{\labelwidth}{1.5em}
    \setlength{\labelsep}{0.5em} } }

\newcommand{\squishend}{
  \end{list}  }

\newcommand{\calG}{\ensuremath{\mathcal{G}}}
\newcommand{\calH}{\ensuremath{\mathcal{H}}}

\begin{document}

% ****************** TITLE ****************************************

\title{Data Placement and Replica Selection for Improving Co-location in Distributed Environments}

% ****************** AUTHORS **************************************

% You need the command \numberofauthors to handle the 'placement
% and alignment' of the authors beneath the title.
%
% For aesthetic reasons, we recommend 'three authors at a time'
% i.e. three 'name/affiliation blocks' be placed beneath the title.
%
% NOTE: You are NOT restricted in how many 'rows' of
% "name/affiliations" may appear. We just ask that you restrict
% the number of 'columns' to three.
%
% Because of the available 'opening page real-estate'
% we ask you to refrain from putting more than six authors
% (two rows with three columns) beneath the article title.
% More than six makes the first-page appear very cluttered indeed.
%
% Use the \alignauthor commands to handle the names
% and affiliations for an 'aesthetic maximum' of six authors.
% Add names, affiliations, addresses for the seventh etc. author(s) as the argument for the
% \additionalauthors command.
% These 'additional authors' will be output/set for you
% without further effort on your part as the last section in
% the body of your article BEFORE References or any Appendices.

%\numberofauthors{4} %  in this sample file, there are a *total*
% of EIGHT authors. SIX appear on the 'first-page' (for formatting
% reasons) and the remaining two appear in the \additionalauthors section.

\author{
\large
\begin{tabular}{ccccc}
  K.~Ashwin Kumar& \mbox{\ \ } & Amol Deshpande & \mbox{\ \ } & Samir Khuller\\[3pt]
\multicolumn{5}{c}{\{ashwin, amol, samir\}@cs.umd.edu} \\[3pt]
\multicolumn{5}{c}{University of Maryland at College Park} \\
\end{tabular}
}

% There's nothing stopping you putting the seventh, eighth, etc.
% author on the opening page (as the 'third row') but we ask,
% for aesthetic reasons that you place these 'additional authors'
% in the \additional authors block, viz.
% Just remember to make sure that the TOTAL number of authors
% is the number that will appear on the first page PLUS the
% number that will appear in the \additionalauthors section.

\maketitle
\begin{abstract}
%by simulating and replaying TPC-H type of queries and our 
%experiments show significant savings in energy consumption.

\eat{Exponential data growth and ever increasing service demands have led to an increase in the number of Web-based services. Scalability has become a primary concern to manage and serve such huge amounts of data.} \eat{Scale-out architectures have emerged as most promising solution to handle the problem of data explosion.}

Increasing need for large-scale data analytics in a number of application domains has led
to a dramatic rise in the number of distributed data management systems, both parallel relational
databases, and systems that support alternative frameworks like MapReduce. There is thus an increasing
contention on scarce data center resources like network bandwidth (especially cross-rack bandwidth); further, the energy
requirements for powering the computing equipment are also growing dramatically. 
As we show empirically, increasing the execution parallelism by spreading out data across a large number of machines may achieve
the intended goal of decreasing query latencies, but in most cases, may increase the total resource and energy consumption 
significantly. For many analytical workloads, however, minimizing query latencies is often not critical; in such scenarios,
we argue that we should instead focus on minimizing the {\em average query span}, i.e., the average number of machines 
that are involved in processing of a query, through co-location of data items that are frequently accessed together.
In this work, we exploit the fact that most distributed environments need to use {\em replication} for fault
tolerance, and we devise {\em workload-driven} replica selection and placement algorithms that attempt
to minimize the average query span. We model a historical query workload trace as a
{\em hypergraph }
%we show experimentally that, in most scenarios, this 
%directly reduces the total resource consumption of
%the query and thus the total energy consumed during execution. We model the query workload
%as a hypergraph 
over a set of data items (which could be relation partitions, or file chunks),
and formulate and analyze the problem of replica placement by drawing connections to several
well-studied graph theoretic concepts. We use these connections to develop a series of algorithms
to decide which data items to replicate, and where to place the replicas.  We 
show effectiveness of our proposed approach by building a trace-driven simulation framework and by presenting results on a collection
of synthetic and real workloads. %\hilight{We further substantiate our results} by experimental evaluation on Amazon EC2 using the standard TPC-H benchmark.\eat{ to show effectiveness of our proposed approach.}  
Our experiments show that careful data placement and replication can dramatically reduce the average query spans resulting in significant reductions in the resource consumption.

\eat{We evaluate our proposed 
techniques by evaluating query span effectiveness by building a trace-driven simulation framework and and also conducting experiments on Amazon EC2 instances on a standard TPC-H benchmark to show effectiveness of our proposed approach. Our experiments show that careful data placement and replication 
can dramatically reduce the average query spans.}
\end{abstract}
\section{Introduction}
\label{sec:intro-expts}

% to provide the most effective solution to this huge problem of energy efficiency in data centers. Therefore, we must design new algorithms and techniques for energy consumption minimization during processing while leveraging the advantages provided by latest energy-friendly hardware.
%These also provide us with an area of opportunity for minimizing the energy consumption. 
%\iffalse
%aug28

%Optimizing the hardware or using energy-friendly hardware can be the one of the 
%ways to achieve energy efficiency. 
%However, this may not provide the complete solution unless we tailor supporting software 
%techniques such as energy efficient algorithms or data structures that compliment these 
%new range of energy-friendly hardware to provide the most effective solution to this huge problem of energy efficiency in data centers. 
%Therefore, we must design new algorithms and techniques for energy consumption minimization during processing while leveraging the advantages provided by latest energy-friendly hardware.

%Optimizing the hardware or using energy-friendly hardware can be the one of the ways to achieve energy efficiency. 
%\fi
\eat{In this evolving world of distributed databases,}
 Massive amounts of data are being generated every day in a variety of domains ranging from
 scientific applications to social networks to retail. The stores of data on which modern businesses
 rely are already vast and increasing at an unprecedented pace. Organizations are capturing data at
 deeper levels of detail and keeping more history than before. %Managing all of the
 %data is thus emerging as one of the key challenges. 
 This deluge of data has led
 to a rapidly increasing use of parallel and distributed data management systems like parallel databases and
 MapReduce frameworks like Hadoop, to analyze and gain insights from the data. A variety of complex
 analysis tasks and queries are executed using these data management systems.
 %predictive analytics
 %intelligence to scientific 
 %interesting trends, 
 %queries or tasks are run on these data management systems to identify interesting trends, to 
 %unusual patterns stand out, or verify hypotheses. 
 In parallel databases, the queries typically
 consist of multiple joins, group-bys on multiple attributes, and complex aggregations. On
 Hadoop, the tasks often have similar flavor, with simplest of map-reduce programs being aggregation tasks
 that form the basis of analysis queries. There have also been many attempts to combine the
 scalability of Hadoop and declarative querying abilities of relational databases~\cite{Thusoo,
 Olston:2008:PLN:1376616.1376726}. 

Use of such parallel or distributed frameworks is expected to accelerate in the coming years,
putting further strain on already-scarce resource like compute power, network bandwidth, and energy.
For reducing total execution times, there is a trend towards increasing the execution parallelism 
by spreading out data across a large number of machines. However, this often increases the 
total resource consumption significantly, as we also illustrate empirically below. We argue that, 
for most analytical workloads, minimizing the query\footnote{We use the term {\em query} to denote
both SQL queries and analysis tasks written using map-reduce or analogous frameworks.} latencies may 
not be critically important since the queries are often not run in an interactive mode. Instead, 
we argue that we should aim for reducing the total resource 
consumption by {\em decreasing} the degree of single-query execution parallelism, i.e., by trying to
reduce the number of machines involved in the execution of a query (called {\em query span}).
There are several advantages to doing that:

%The increasing use of such parallel or distributed frameworks 
%
%For fault tolerance, load balancing, and availability, these systems usually keep several copies of each data item (e.g., Hadoop file system (HDFS) maintains at least 3 copies of each data item by default~\cite{citeulike:4882841}). Our goal in this work is to 
%show how to exploit this inherent replication in these systems to minimize the number of machines that are involved in executing a query, called the {\em query span} (we use the term query to denote both SQL queries and Hadoop tasks). 
%There are several motivating reasons for doing this:
%Why do we want to do it? There are many reasons:
\begin{figure*}[t]
\centering
\hspace{-0.16in}
\subfigure[]{
\includegraphics[scale=0.89]{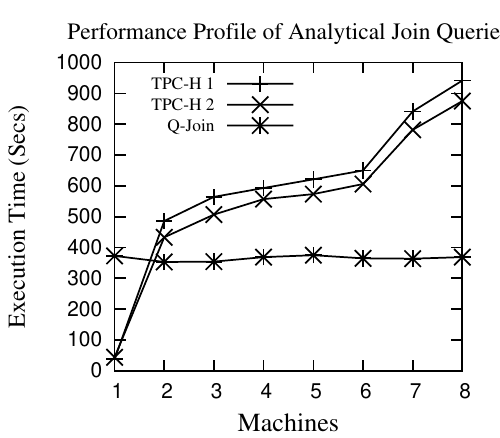}
\label{fig:mot1}
}
\hspace{-0.16in}
\subfigure[]{
\includegraphics[scale=0.89]{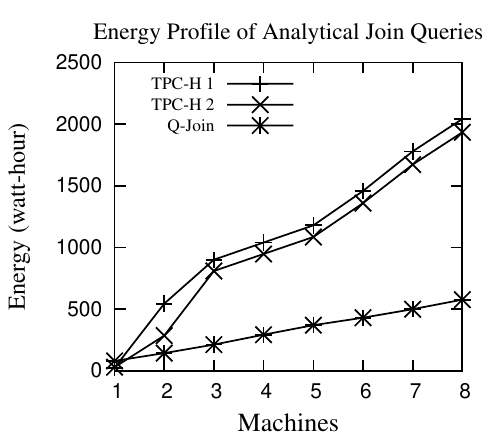}
\label{fig:mot2}
}
\hspace{-0.16in}
\subfigure[]{
\includegraphics[scale=0.89]{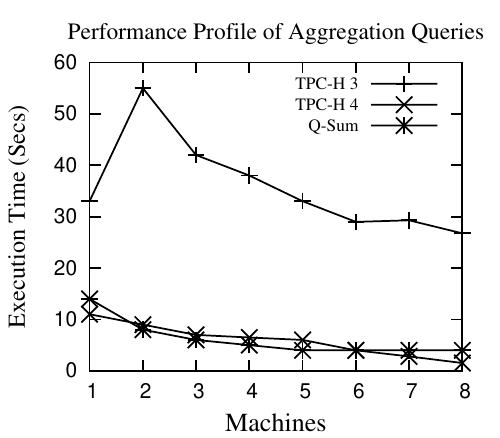}
\label{fig:mot3}
}
\hspace{-0.16in}
\subfigure[]{
\includegraphics[scale=0.89]{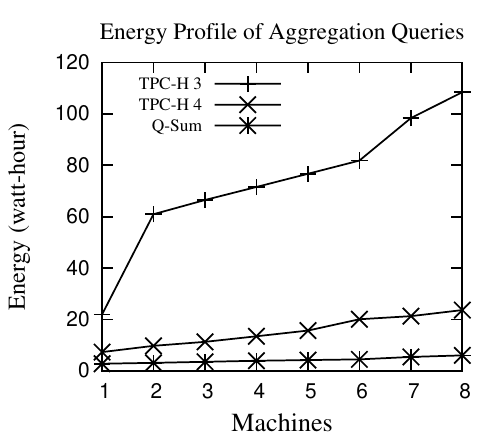}
\label{fig:mot4}
}

\vspace{-.15in}
\caption[]{Increasing the degree of parallelism may result in lower execution times, but leads to significantly 
higher resource and energy consumption.}
\label{fig:motivation}
\vspace{-.15in}
\end{figure*}

\emtopic{Minimize the communication overhead:} Query span directly impacts the total communication that must be performed to execute a query. This is clearly a concern in distributed setups (e.g., grid systems~\cite{grid2-ch19} or multi-datacenter deployments); however even within a data center, communication network is oversubscribed, and especially cross-rack communication bandwidth can be a bottleneck~\cite{10.1109/CLOUD.2011.17, DBLP:conf/sigcomm/ChowdhuryZMJS11}. 
%Another aspect that is often considered is the rack placement of replicas in Hadoop. 
In cloud computing, the total communication directly impacts the total dollar cost of executing a query. 
HDFS, for instance, tries to place all replicas of a data item in a single rack to minimize inter-rack data transfers~\cite{citeulike:4882841}. 
We take this further, and argue for clustering replicas of different data items together to improve
network performance for queries that access multiple data items.
%Our algorithms can be used to further guide these decisions and cluster replicas of multiple data
%items on to a single rack to improve network performance for queries that access multiple data items, which HDFS currently ignores. 

\emtopic{Minimize the total amount of resources consumed:} It is well-known that parallelism comes
with significant startup and coordination overheads, and we typically see sub-linear speedups as a
result of these overheads and data skew~\cite{Pavlo:2009:CAL:1559845.1559865}. Although the response
time of a query usually decreases in a parallel setting, the total amount of resources consumed
typically increases with increased parallelism. Even in scenarios where we obtain {\em
super-linear} speedups due to higher aggregate memory across the machines, we expect the
total energy consumption to increase with the degree of parallelism.

\emtopic{Reduce the energy footprint:} Computing equipment in US costs data center operators millions of dollars
annually for energy, and also impacts the environment. Energy costs are ever increasing and hardware costs are decreasing -- as a result soon the energy costs to operate and cool a data center may exceed the cost of the hardware itself. Minimizing the total amount of resources consumed directly reduces the total energy consumption of the task. 

%\comment{I had to repeat some of the below description}\\
%\comment{ in the experiments section, may be redundant}

\topic{Illustrative Experiments:}
To support these claims and to motivate query span as a key metric to optimize,
we conducted a set of experiments analyzing the effect of query span on the total resource and
energy consumption under a variety of settings. First setting is a horizontally
partitioned MySQL cluster, on which we execute four SQL queries against a TPC-H database. % we evaluate a total of four TPC-H benchmark queries. 
Two of the queries are complex analytical join queries (TPC-H1, TPC-H2 in Figure
\ref{fig:motivation}), whereas the other two are simple aggregation queries on a single table
(TPC-H3, TPC-H4). 
In the second setting, we implemented our own distributed query processor on
the top of multiple MySQL instances running on a cluster where predicate evaluations are pushed on to the individual nodes and data is shipped to 
a single node for perform the final steps.
%master then performs the final processing. 
On this setup we evaluate two queries: a complex join query (Q-Join)
and a simple aggregate query on a single table (Q-Sum). In Figures~\ref{fig:mot1} and \ref{fig:mot2}, we plot the execution times and the energy consumed as the number of machines across
which the tables are partitioned (and hence query span) increases. 
The energy consumption is estimated by using an Itanium server power model
constructed by using the Mantis full-system power modelling technique~\cite{Economou06full-systempower}. 
We use the \emph{dstat} tool to collect various system performance counters such as CPU utilization,
network reads and writes, I/O, and memory footprint, and then use the power
model to estimate the total energy consumed.
%with the power model is used to compute the total energy consumed. %We use

As we can see, the execution times of the TPC-H
queries run on MySQL cluster actually increased with parallelism, which may be because of nested loop join implementation in MySQL cluster (a known problem that is being fixed). 
In our implementation, the execution time remains constant. But in all cases, energy consumption
increased with query span.
In the second experiment with simpler queries (Figures~\ref{fig:mot3} and \ref{fig:mot4}), though execution times decrease as the query span increases, energy consumption increases in all cases. 
%these counters as a proxy for the resources used to execute a query/job. 
From this simple set of experiments it is evident that, as the number of machines involved in processing a query increases, total resources consumed to process the query also rise.

\topic{Goals and Contributions:}
%\vspace{8pt}
%\noindent{In} 
In this paper, we propose a {\em workload-driven} approach that aims to reduce the average 
query span in distributed data management systems by 
co-locating data items that are frequently accessed together by queries. We observe
that, for fault tolerance, load balancing, and availability, those systems
typically maintain several copies of each data item (e.g., Hadoop file system (HDFS) maintains at
least 3 copies of each data item by default~\cite{citeulike:4882841}), and we propose exploiting 
this inherent replication to achieve higher co-location by judicious replica creation and
placement.  Our approach is workload-driven in that, we propose capturing a historical query
workload over a period of time, and optimizing data placement and replication for that workload. 
%We propose exploiting this inherent 
%Our goal in this work is to 
%replication 
Our techniques work on an abstract representation of the query workload, and are 
applicable to both multi-site data warehouses and general purpose data centers. %We assume that a query workload trace
%is provided that lists the data items that need to be accessed to answer each query. 
We represent the query workload as a {\em hypergraph}, where the nodes are the data items and each 
query is translated into a {\em hyperedge} over the nodes.  The data items
could be database relations, parts of database relations (e.g., tuples or columns), or arbitrary files. 
The goal is to store each data item (node in the graph) onto a subset of machines/sites (also called {\em partitions}), obeying the storage capacity requirements
for the partitions. Note that the partitions do not have to be machines, but could instead represent
racks or even datacenters. %This specifies the layout completely. 
The span of a query is defined to
be the smallest number of partitions that contain all the data that the query needs. Our goal is to
find a layout that minimizes the average span over all queries in the workload. Further, our algorithms can optimize for load or storage constraints, or both.

%In this paper, we address the problem of minimizing the average query span for a query workload through judicious replica selection (by choosing which data items to replicate and how many times), and data placement. 

%Our key contributions are the following:
%\squishlist
%\item \emph{Problem formulation and analysis:}
%\item \emph{Problem formulation and analysis:}
%\squishend

%\comment{Can we make below paragraph more concrete?}

Our key contributions include formulating and analyzing this problem, drawing connections to 
several problems studied in the graph algorithms literature, and developing efficient algorithms 
for data placement. In addition, we examine the special case when each query accesses at most
two data items -- in this case the hypergraph is simply a graph. For this case, we are able to 
develop theoretical bounds for special classes of graphs that gives an understanding
of the trade-off between energy cost and storage. 
We have also built a trace-driven simulation framework that enables us to systematically compare different
algorithms, by automatically generating varying types of query workloads and by calculating the 
total energy cost of a query trace. We conducted an extensive experimental evaluation using
our framework, and our results show that our techniques can result in high reductions in query spans and resource
consumption compared to baseline or random data placement approaches. 
%\hilight{that can help minimize distributed overheads}. \\
%\comment{not clear, depends on the query, isn't it?}\\
%\comment{or do we just mean resource consumption?}

\topic{Discussion:}
Making data placement and replication decisions with the goal of minimizing average query spans
raises several concerns. 
First, as discussed above, it may increase the execution time of a single query, and hence such an
approach can only be used if the workload is not latency-sensitive. We argue that an increasing
number of analytical workloads, especially those primarily consisting of batch analysis tasks, fall
in that category. If the primary goal is to minimize the query response time, then {\em
declustering} should instead be utilized to leverage the parallelism by spreading out the data items.
Second, focusing simply on minimizing query spans can lead to a
load imbalance across the partitions. There are two ways this could be handled. We can use 
temporal scheduling (by postponing certain queries) to balance loads across machines. 
%believe total resource consumption should be the key optimization goal in most situations. Most 
%analytical workloads are typically not latency-sensitive, and we can use temporal scheduling 
%(by postponing certain queries) to balance loads across machines. 
We can also easily modify our algorithms to incorporate load constraints. 
A third concern is the cost of replica maintenance. However, most distributed
systems do replication for fault tolerance, and hence our approach does not add a significant extra 
overhead. Further, most systems geared towards large-scale analytics perform batch updates, and the
overall cost of updates is relatively low. Finally, like any workload-driven approach, our proposed
approach relies on the ability to capture and model an expected query workload. With increasing
automation in data analysis, with the same queries or analysis tasks being run on a regular basis,
we believe this is a reasonable assumption to make.

Our proposed techniques have broader applicability beyond the application domains that we discuss in
this paper. We can use similar techniques to partition large graphs across a distributed cluster; smart
replication of some of the (boundary) nodes can result in significant savings in the communication cost
to answer queries~\cite{sigmod2012jayanta}. %(e.g., to answer subgraph pattern queries\eat{~\cite{matthias}}). 
Our techniques are also applicable in partition farms such as MAID~\cite{762819}, PDC~\cite{1006220},
or Rabbit~\cite{socc2010}, that utilize a subset of a partition array as a workhorse to store popular data so that other partitions could
be turned off or sent to lower energy modes. A recent system, CoHadoop~\cite{DBLP:journals/pvldb/EltabakhTOGKM11}, also aims at 
co-locating related data items to improve performance of Hadoop, and provides a lightweight mechanism that allows applications to control where data is
stored. They focus on data co-location to improve the efficiency of many operations, including indexing, grouping, aggregation, columnar storage, joins, 
and sessionization. Our workload-driven techniques are complimentary to their work, and can be 
used to further guide the data placement decisions in their system. 

In a recent work, Curino et al.~\cite{DBLP:journals/pvldb/CurinoZJM10} also proposed a workload-aware approach
for database partitioning and replication to minimize the number of sites involved in distributed
transactions. They however do not develop new partitioning techniques. Although there are superficial similarities 
in use of graph partitioning techniques, there are several major
differences. First, the number of data items is significantly higher in that application domain (since the approach treats
each tuple as a data item); second, we largely assume a read-only workload, but in their setting, replication costs 
must be taken into account. We note that, in a concurrent submission by a subset of the authors~\cite{abdulsubmission},
we propose a suite of techniques for scalable workload-aware data partitioning and replication for OLTP workloads, that
builds upon the work by Curino et al. Unlike this submission where the focus is on developing new partitioning and replication
algorithms, in that work, our focus is on minimizing the partitioning and bookkeeping overheads, on minimizing update costs through
use of quorums, and on handling dynamic changes to the workload through incremental re-partitioning.

\eat{Our algorithms can be applied to minimize overheads at various different granularities in a data center. For
example, in shared nothing scenarios, we can use similar techniques to reduce the number of active
servers at any time. Networking infrastructure including switches and routers are also known to be
highly energy-inefficient~\cite{5507749} and cross switch/router communications are expensive, and we can use judicious data placement to improve data locality and reduce cross switch communication.}

%We note that significant work has been done on the converse problem of minimizing query response times or latencies. 
%{\em Declustering} refers to the approach of leveraging parallelism in the partition subsystem by spreading
%out blocks across different partitions so that multi-block requests can be executed in parallel.
%%different partitions and thus entail a parallel execution of the request. 
%In contrast, we try to cluster data items together to minimize the number of sites 
%required to satisfy a complex analytical query. 

\topic{Outline:} We begin with a discussion of closely related work (Section \ref{sec:related
work}). \eat{We then analyze the key components of partition energy consumption (Section \ref{sec:partitions}, and}We formally 
define the problem that we address in the paper and analyze it (Section \ref{sec:definition}). We
present a series of algorithms to solve the problem (Section \ref{sec:algorithms}), and present an 
extensive performance evaluation using real dataset on Amazon EC2 and trace-driven simulation framework that we have built
(Section \ref{sec:expts}). 
%Finally, we present a brief discussion Section \ref{sec:discussion}) on effect of our proposed approach on query response times and resource consumption by showing experimentation result on TPC-H queries.

\section{Related Work}
\label{sec:related work}

Data partitioning and replication plays an increasingly important role in large scale distributed
networks such as content delivery networks (CDN), distributed databases and distributed systems such
as peer-to-peer networks. Due to space constraints, we limit our discussion to the most relevant
work here, and refer to the extended version for further discussion~\cite{fullpaper}. 
%Recent work~\cite{Dittrich:2010:HMY:1920841.1920908, Abouzeid:2009:HAH:1687627.1687731} has shown that judicious placement of data and replication
%improves the efficiency of query processing algorithms. 
Aside from CoHadoop work discussed above~\cite{DBLP:journals/pvldb/EltabakhTOGKM11}, 
%There has been some recent interest on improving data co-location in large scale processing systems like Hadoop. Recent work by
    %Eltabakh et al.~\cite{DBLP:journals/pvldb/EltabakhTOGKM11} on CoHadoop is very close to our
    %work, where they provide an extension for Hadoop with a lightweight mechanism that allows
    %applications to control where data is stored. They focus on data co-location to improve the
    %efficiency of many operations, including indexing, grouping, aggregation, columnar storage,
    %joins, and sessionization. Our techniques are complimentary to their work. %, where our
    %algorithms can help them to perform aggressive data co-location while placing the replicas. 
Hadoop++~\cite{Dittrich:2010:HMY:1920841.1920908} is another closely related work that exploits data
pre-partitioning and co-location. There is substantial amount of work on replica placement that
focuses on minimization of network latency and bandwidth. Neves et
al.~\cite{journals/endm/NevesDOAU10} propose a technique for replication in CDN where they replicate
data onto a subset of servers to handle requests so that the traffic cost in the network is
minimized. There has been a lot of work on dynamic/adaptive replica
management (e.g.,~\cite{Wolfson97anadaptive}), %, springerlink:10.1007/3-540-45644-9_8,
%10.1109/CCGRID.2002.1017164, 1281601, springerlink:10.1007/s11227-009-0371-9,
%Zhang:2009:NRP:1506477.1506511}, 
    where replicas are dynamically placed, moved, or deleted based on the read/write access frequencies of the data items again with the goal of minimizing bandwidth and access latency. 
%Our work is complimentary to this line of work, in a way that we replicate the data considering the query workload nature by modelling it as hypergraph. Extending our approach to track changes in the query workload and adapt the replication decisions is an interesting direction for future work that we are planning to pursue.

Graphs have been used as a tool to model various distributed storage problems and to come up with
replication strategies to achieve a specific objective. Du et al.~\cite{Du20111224} study
Quality-of-Service (QoS)-aware replica placement problem in a general graph model. In their model,
vertices are the servers with various weights representing node characteristics and edges
representing the communication costs. Other work has modeled network topology as a graph and
developed replication strategies or approximations (replica placement in general graphs is
NP-complete)~\cite{Wolfson:1991:MPR:103140.103146}. In contrast, %This is different from what we are doing in this
%paper: 
we model query workload as a hypergraph, and %, whereas that prior work models network topology as graph. 
assume a uniform network topology (i.e., identical communication costs between any pair of nodes);
we believe this better approximates the current networks. %As discussed earlier, Curino

Issues in energy-efficient computing are being increasingly studied at all layers of 
today's computing infrastructures. %We briefly discuss the related work here.
%energy efficiency by designing low-power hardware~\cite{jhamilton,szalay} and energy-proportional
%hardware (where the energy consumed is proportional to the amount of work done)~\cite{citeulike:3888154}. 
%Software-level optimizations are also being studied in many domains. %, e.g., sensor
%networks (where battery power is often a significant resource constraint). 
%However, there isn't
%much work in this area within the database community.
%Increasingly there is also work on software-level optimizait
%and energy-proportional hardware
%With energy costs forming an increasing fraction of the total cost of computing infrastructure, 
%energy-efficient computing
%Energy efficient computing~\cite{citeulike:3888154,belady,kgbrill,citeulike:1397440,jhamilton,jgkoomey}
%is relatively a new topic for the database community compared to sensor networks and other areas where
%energy constraints are a concern. Hence, the whole new exciting field is open for database
%researchers. 
Harizopoulos et al.~\cite{DBLP:journals/corr/abs-0909-1784} reported the first results on
software-level optimizations to achieve better energy efficiency; they experiment with a system that
was configured similarly to an audited TPC-H server and show that making the right physical design decisions 
can improve energy efficiency. Additionally, they use relational scan operator as a
basis to demonstrate that optimizing for performance is different from optimizing for energy
efficiency. It is also among the first papers~\cite{DBLP:journals/corr/abs-0909-1784,1385494,DBLP:conf/cidr/LangP09} to 
practically show the importance of energy efficiency in database systems. %Graefe
%\cite{DBLP:journals/corr/abs-0909-1784} also points out various research challenges and promising
%approaches in
%energy-efficient database management. 
%In his paper he indicates various promising approaches and techniques to achieve energy efficiency in database systems. Lang et al.~\cite{DBLP:conf/cidr/LangP09} examine techniques to trade energy for performance
%in distributed computing environments.
%again in the context of database systems. They basically investigate energy efficient methods for
%general data processing in distributed computing environments. 
%He discusses two approaches in this context: processor frequency control and explicit delays. 
Leverich et al.~\cite{1740405} and Lang et al.~\cite{DBLP:journals/pvldb/LangP10} suggest
approaches to conserving energy by powering down Hadoop cluster nodes. 
%during periods of low load, and observe that the default replica placement policy is highly 
%inefficient in this regard. In particular, they observe that powering down any three nodes 
%is likely to lead to some data being unavailable, and instead suggest a replication 
%policy such that a small set of cluster nodes {\em cover} (contain) at least one replica of
%each data item. 
%%do something very similar where they powers down MapReduce cluster nodes during periods of low load, and we give better solutions for that. 
%Lang et al.~\cite{DBLP:journals/pvldb/LangP10} suggest and evaluate an alternative approach where 
%all cluster nodes are powered up (to answer queries), and powered down at the same time, and show that their approach leads
%to better energy utilization. The approach that we consider here is more fine-grained in that, we consider shutting down 
%individual disks (or nodes) during periods of low load, and wake them up as needed.
Tsirogiannis et al.~\cite{1807194} analyze the energy efficiency of a single-node database
server, and argue that the most energy-efficient configuration is typically the highest
performing one. However, this assertion is valid only for single node database server, and does not
hold for scale-out architectures involving multiple machines where parallelization, communication, and startup
overheads come into play. %Our work provides one such result where we consider these overheads.
From our experiments over the TPC-H benchmark, it is evident that, as the number of machines involved in
processing a query increases, total resources consumed to process the query also rise. %We also show
%that for our setup this result holds true irrespective of the nature of the query. We do not
%consider situations where we get super linear speedups on parallelization.

\eat{We note that this does not contradict our results here, since we consider read only TPC-H style of queries where most of the queries have complex aggregation and multiple join conditions, our approach does not necessarily lead to higher query response times, and in fact, by reducing the number of nodes involved in executing the query, it leads to better query response times. Our initial experiments bolster this claim. FAWN~\cite{1629577} focuses on scale-out architecture where each node is energy efficient delivering more queries per Joule.}
%Following their result, it is worth noting that in our approach due to data locality and replication, query response time doesn't get worse, in fact it may improve by avoiding extra query result joins and network delays. We draw various insights from these initial works in energy-efficient database systems, and try to solve a problem of minimizing average energy consumed by a database query with respect to storage subsystem, by partitioning and replicating the data and placing it judiciously on an array of partitions.

Our work is different from several other works on data placement~\cite{koyuturk05,DBLP:journals/is/LiuS96,1616815} 
where the database query workload is also modeled as a hypergraph and partitioning techniques are used to 
drive data placement decisions. 
%Liu et al.~\cite{DBLP:journals/is/LiuS96} propose a novel declustering technique based on max-cut partitioning of a weighted similarity graph. 
%Aykanat et al.~\cite{koyuturk05} observe that the approach where each query over a set of relations
%is represented by a clique over those relations, does not accurately capture the cost function, and instead
%propose directly using the hypergraph representation of the query workload. 
Tosun et al.~\cite{Tosun97optimalparallel,968054} and Ferhatosmanoglu et al.~\cite{1055577} propose
using replication along with declustering for achieving optimal parallel I/O for spatial range queries.
The goal with all of that prior work is typically minimization of latencies and
query response times by spreading out the work over a large number of partitions or devices. 
%Under the framework that we consider here, this is exactly the wrong optimization goal -- we would 
For us, that is exactly the wrong optimization goal -- we would 
like to cluster data required for each query on as few partitions as possible.

The problems we study are closely related to several well-studied problems in graph theory and can be considered 
generalizations of those problems. A basic special case of our main problem is the {\em minimum graph bisection}
problem (which is NP-Hard), where the goal is to partition the input graph into two equal sized partitions, while 
minimizing the number of edges that are cut~\cite{boppana}. There is much work on both that problem and its
generalization to hypergraphs and to $k$-way partitioning~\cite{1577032,Karypis99multilevelhypergraph,Karypis98multilevelk-way}. 
%The work on {\em community detection} over complex networks~\cite{fortunato} has also proposed many schemes for
%partitioning graphs to minimize the connections between partitions; however the resulting partitions there do not
%have to be balanced -- a critical requirement for us.
Another closely related problem is that of finding {\em dense subgraphs} in a graph, where the goal is to find a group of vertices where
the number of edges in the induced subgraph is maximized~\cite{Feige99thedense}. 
Finally, there is much work 
on finding {\em small} separators in graphs. %i.e., small groups of vertices that separate the graph into almost
%%equal-sized partitions (this is different from graph bisection, since the goal there is to find a small group
%of edges)~\cite{966128, 500866}. 
Several theoretical results on known about this problem. We discuss these 
connections in more detail later when we describe our proposed algorithms.

\eat{There is also much work on the problem of accurately modeling partition energy consumption~\cite{mhd,DBLP:conf/mascots/HylickSRJ08,Greenawalt94modelingpower}.
Hylick et al.~\cite{hylick} develop a simple methodology 
for creating accurate hard drive runtime energy models
through the use of easily obtainable data derived from published specifications and performance
measurements. %Their work is important because of simplicity and accuracy of their partition power
%estimation model. One can easily estimate the power components of the partition drive using the published
%partition power and performance specifications. 
We directly use their model to estimate partition drive energy
consumption in our trace-driven simulation framework.}

\section{Problem Definition; Analysis}
%\label{sec:problem}
\label{sec:definition}
Next, we formally define the problem that we study, and draw connections to some closely 
related prior work on graph algorithms. 
We also analyze a special case of the problem formally,
and show an interesting theoretical result.

\topic{Problem Definition:}
Given a set of data items $\mathcal{D}$ and a set of partitions, our goal is to decide which data items to replicate
and how to place them on the partitions to minimize the average span of an expected
query workload; {\em span} of a query is defined to be the minimum number of partitions that must be accessed
to answer the query. To make the problem more concrete, we assume that we are given a set of queries over the 
data items, and our goal is to minimize the average span over these queries. For simplicity, we assume that we
 are given a total of $N$ identical partitions each with capacity $C$ units, and further that the data items are 
all unit-sized (we will relax this assumption later). Clearly, the number of data items must be smaller than
$N\times C$ (so that each data item can be placed on at least one partition). Further, let $N_e$ denote the minimum 
number of partitions needed to place the data items (i.e., $N_e = \lceil |\mathcal{D}|/C \rceil$).

\begin{figure}
    \centering 
    \includegraphics[width=3.5in]{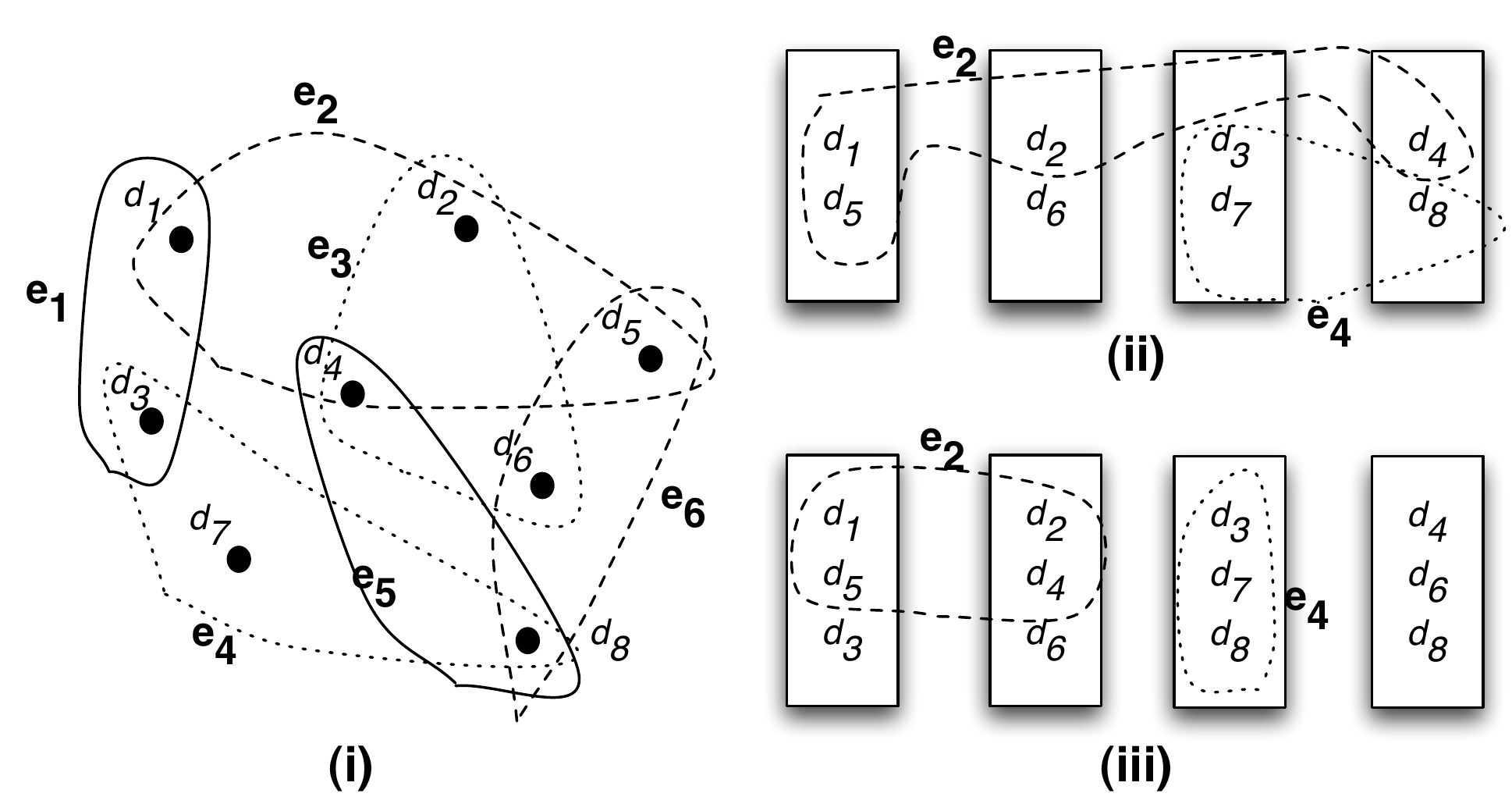} 
    \caption{(i) Modeling a query workload as a hypergraph -- $d_i$ denotes the data items, and $e_i$ denotes the queries represented
    as hyperedges; (ii) A layout w/o replication onto 4 partitions -- the span of two of the hyperedges is also shown; (iii) A layout with
    replication -- span for both queries reduces by 1.}
    \label{fig1} 
\end{figure} 

The query workload can be represented as a hypergraph, $\mathcal{H} = (V, E)$, where the nodes are the data items and
each (hyper)edge $e \in E$ corresponds to a query in the workload. Figure \ref{fig1} shows an illustrative example, 
where we have 6 queries over 8 data items, each of which is represented as a hyperedge over the data items.
%$Q=\{e1: \{d_1, b, c, d, e\},$\\$q2: \{b, c, d, e, f\}, q3: \{c, d, e, f, g\}, q4: \{d, e, f, g, h\}, q5: \{e, f, g, h, i\}, q6: \{f, g, h, i, j\}, q7: \{g, h, i, j, k\}, q8: \{h, i, j, k, l\}\}$. 
The figure also shows two layouts
of the data items onto 4 partitions of capacity 3 each, without replication and with replication.

\topic{Calculating Span:} When there is no replication, calculating the span of a query is straightforward since each data item
is associated with a single partition. However, if there
is replication, the problem becomes NP-Hard. It is essentially identical to the {\em minimum set cover} problem~\cite{garey-johnson:79}, where we
are given a collection of subsets of a set (in our case, the partitions) and a query subset, and we are asked to find the minimum
number of subsets (partitions) required to cover the query subset. 

As an example, for query $e_2$ in Figure \ref{fig1}, the span in the first layout is 3. However, in the second layout, 
we have to choose which of the two copies of $d_4$ to use for the query. Using the first copy (on second partition) leads to 
the lowest span of 2.
Overall, the average query span for the first layout is $13 \over 6$, but use of replication in the second layout reduces this to $8 \over 6$.

We use a standard greedy algorithm for choosing replicas to use for a query and for calculating the span. 
For each of the partitions, we compute the size of its intersection with the query subset. We choose the partition 
with the highest intersection size, remove all items from the query subset that are contained in the partition, and 
iterate until there are no items left in the query subset.
This simple greedy algorithm provides the best known approximation to the set cover problem ($\log |Q|$, where $|Q|$ is the query size).
%In the rest of the paper, we use this algorithm for calculating the span.

\topic{Hypergraph Partitioning:}
Without replication, the problem we defined above is essentially the $k$-way 
(balanced) hypergraph partitioning problem that has been very well-studied
in the literature. 
However, the optimization goal of minimizing the average span is 
unique to this setting; prior work has typically
studied how to minimize the number of {\em cut} hyperedges instead. 
%XXX Add a discussion on the objective function from HMetis that is the closest.
Several packages are available for partitioning very large hypergraphs efficiently~\cite{hMETIS,MLPart}. The proposed algorithms are typically heuristics or 
combinations of heuristics, and most often the source code is not available. 
We use one such package (hMETIS) as the basis of our algorithms.

%However, the problem with data replication is not 
%well-studied in the literature.
%XXX discuss this somewhat, and propose some heuristics.

\topic{Finding Dense Subgraphs of a specified size:}
Given a set of nodes $S$ in a graph, the {\em density} of the subgraph induced by $S$ is defined to be the 
ratio of the number of edges in the induced subgraph and $|S|$. The dense subgraph problem is to find the densest
subgraph of a given size.
To understand the connection to the dense subgraph problem, consider a scenario where we have exactly one ``extra'' partition for replicating
the data items (i.e., $N_e = N - 1$). 
Further, assume that each query refers to exactly two data items, 
i.e., the hypergraph $\mathcal{H}$ is just a graph. 
One approach would then be to first partition the data items into 
$N-1$ partitions without replication, and then try to use
this extra partition optimally. To do this, we can construct a {\em residual} graph, which contains all edges that were cut in this 
partitioning. The span of each of the queries corresponding to these edges is exactly 2. Now, we find the subgraph of size $C$
such that the number of induced edges (among the nodes of the subgraph) 
is maximized, and we place these data items on the 
extra partition. The span of the queries corresponding to these edges are all reduced from 2 to 1, and hence this is an optimal 
way to utilize the extra partition. We can generalize this intuition to hypergraphs and this forms the basis of one of our 
algorithms. 

Unfortunately, the problem of finding the most dense subgraph of a specified 
size is NP-Hard (with no good  
worst case approximation guarantees), so we have to resort to heuristics. 
One such heuristic that we adapt in our work is as follows: 
recursively remove the lowest degree node from the residual graph (and all its
incident edges) till the size of the residual graph is exactly $C$. 
This heuristic has been analysed by Asahiro et al.~\cite{asahiro:swat96} who find that this simple greedy algorithm can solve this problem with approximation ratio of 
approximately $2(\frac{|V|}{C}-1)$ (when $C \le |V|/3$).

\topic{Sublinear Separators in Graphs:}
%\label{topic:sublinear}
Consider the special case where $\mathcal{H}$ is a graph, and further assume that there are only 2 partitions (i.e., $N = 2$). 
Further, lets say that the graph has a small {\em separator}, i.e., a set of nodes whose deletion results in two connected
components of size at most $n/2$. In that case, we can replicate the separator nodes (assuming there is enough redundancy) and 
thus guarantee that each query has span exactly 1. The key here is the existence of small separators of bounded sizes. Such  
separators are known to exist for many classes of graphs, e.g., for any family of graphs that excludes 
a minor~\cite{DBLP:conf/stoc/AlonST90}.

A separator theorem is usually of the form that, any $n$-vertex graph can be partitioned into two sets $A$, $B$, such that $|A \cap B| = c\sqrt{n}$ for 
some constant $c$, $|A - B| < 2n/3$, $|B - A| < 2n/3$, and there are no edges from a node in $A - B$ to a node in $B - A$. This directly suggests an algorithm that recursively 
applies the separator theorem to find a partitioning of the graph into as many pieces as required, replicating the separator nodes to 
minimize the average span. Such an algorithm is unlikely to be feasible in practice, but may be used to obtain theoretical bounds or approximation algorithms. For example, we prove that:

\vspace{-2pt}
\begin{theorem}
Let $G$ be a graph with $n$ nodes that excludes a minor of constant size. Further, let $N_e$ denote the number of partitions minimally required
to hold the nodes of $G$ (i.e., $N_e = \lceil n/C \rceil$). Then, asymptotically, $N_e^{1.73}$ partitions are enough to partition the nodes of $G$ with 
replication so that each edge is contained completely in at least one partition.
\end{theorem}

\vspace{2pt}
%For general graphs, in Appendix~\ref{appendixb}, we show that:
For general graphs, we show that:
\begin{theorem}
If the optimal solution uses $\beta N_e$ partitions to place the data items so that each edge is contained in at least
one partition, %where $N_e=\lceil n/C \rceil$,
then either we can get an approximation with factor $\frac{2}{2-\alpha}$
for $0 \le \alpha \le 1$ using $N_e$ partitions, or a placement 
using $\frac{C N_e\beta}{2 \alpha}$ partitions with span 1 for each edge.
\end{theorem}

Proofs of both the theorems can be found in the extended version of the paper~\cite{fullpaper}.

\section{Data Placement Algorithms}
\label{sec:algorithms}
In this section, we present several algorithms for data placement with replication, with the 
goal to minimize the average query span. Instead of starting from scratch, we chose to base our
algorithms on existing hypergraph partitioning packages. As we discussed in the previous sections,
the problem of balanced and unbalanced hypergraph partitioning has received a tremendous amount
of attention in various communities, especially the VLSI community. Several very good packages are freely available for 
solving large partitioning problems~\cite{hMETIS, Karypis99multilevelhypergraph, MLPart, 384247}. We use a hypergraph partitioning 
algorithm (called HPA) as a blackbox in our algorithms, and focus on replicating data items appropriately to reduce the 
average query span. An HPA algorithm typically tries to find a balanced partitioning (i.e., all partitions
are of approximately equal size) that minimizes some optimization goal. Usually, allowing for unbalanced 
partitions results in better partitioning. In the algorithm descriptions below, we assume that 
the HPA algorithm can return an exactly balanced partition, where all partitions are of equal size, 
if needed. % (the actual algorithm we use in our experiments, hMETIS, does not have this property -- we modify the returned
%partitioning slightly to enforce it).

Following the discussion in the previous section, we develop four classes of algorithms:

%Hypergraph partitioning problem has been looked into by researchers since a decade. Many efficient algorithms have been developed to partition the hypergraph into K-partitions both balanced and unbalanced whereas hypergraph partitioning with replication has received a very little attention. Given any hypergraph partitioning algorithm $HPA$, we mainly develop algorithms that take the partitions output by $HPA$ and replicate the nodes to improve the partition quality or to maximize the number of local hyperedges captured by each partition. We consider homogeneous data items with equal data sizes. We identify three types of algorithms to improve the partition solution returned by $HPA$ by replicating the nodes,
%\begin{itemize}
%\squishlist
\begin{list}{$\bullet$}{\leftmargin 0.15in \topsep 1pt \itemsep 0pt}
\item \textbf{Iterative HPA (IHPA)}: Here we repeatedly use HPA until all the extra space is utilized.
\item \textbf{Dense Subgraph-based (DS)}: Here we use a dense subgraph finding algorithm to 
    utilize the redundancy.
\item \textbf{Pre-replication (PR)}: Here we attempt to identify a set of nodes to replicate a priori, modify the 
    input graph by replicating those nodes, and then run HPA to get a final placement.
\item \textbf{Local Move-based (LM)}: Starting with a partition returned by HPA, we improve it by 
    replicating a small group of data items at a time.
\end{list}
%\squishend

\noindent{}As expected the space of different variants of the above algorithms is very large. We experimented
with many such variants in our work. We begin with a brief listing of some of the key subroutines that we use in the pseudocodes.
We then describe a representative set of algorithms that we use in our performance evaluation.
%our performance evaluation. 

%\newpage
\subsection{Preliminaries; Subroutines}
%In this section, we provide the subroutines of the algorithms that we develop in this paper. 
The inputs to the data placement algorithm are: (1) the hypergraph, $\mathcal{H}(V, E)$, with vertex set $V$ and (hyper)edge set $E$ that captures the query workload, and (2) the number of partitions, $N$ and (3) the capacity of each partition $C$. We use $N_e$ to denote the minimum number of partitions needed to partition the hypergraph ($N_e \le N$).

Our algorithms use a hypergraph partitioning algorithm (HPA) as a blackbox. HPA takes as input
the hypergraph to be partitioned, the number of partitions, and an \textit{unbalance factor} (UBfactor). The 
unbalance factor is set so that HPA has the maximum freedom, but 
the number of nodes placed in any partition does not exceed $C$. For instance, if $|V| = N_e \times C$ and
if HPA is asked to partition into $N_e$ partitions, then the unbalance factor is set to be the minimum. However, if HPA
is called with $N' > N_e$ partitions, then we appropriately set the unbalance factor to the maximum possible. The formula we use in our experiments to set unbalance factor is: 

\vspace{-6pt}
{\small
\begin{equation*}
UBfactor = 100 * \frac{partitionCapacity * noPartitions - totalItems}{totalItems * noPartitions}
\label{ubfactor}
\end{equation*}
}

\vspace{-4pt}
%XXX Maybe
%some more details and the formula here.
We modify the output of HPA slightly to ensure that the partition capacity
constraints are not violated. This is done as follows: if there is a partition that has higher than maximum number of nodes, we move a small group of nodes to another partition with fewer than maximum number of nodes. 
We use one of our algorithms developed below (LMBR) for this purpose.
%;LMBR technique can be used for this purpose.
% -- in some cases, HPA may place more than $C$ nodes in a partition.

In the pseudocodes shown, apart from HPA, we also assume existence of the following subroutines:
\begin{list}{$\bullet$}{\leftmargin 0.15in \topsep 1pt \itemsep 0pt}
\item {\bf avgDataItemsPerQuery$(\mathcal{H})$: } Suppose $V_i$ is the set of data items covered by
hyperedge $e_i\in \mathcal{H}$. The $\Sigma_{e_i\in \mathcal{H}} |V_i|$ gives the average number of data items covered per query.
    \item {\bf getSpanningPartitions($\calG, e$):} Let the current placement (during the course of the algorithm) be $\calG = \{G_1, \cdots, G_N \}$ where $G_1, \cdots, G_n$ denote 
        the subgraphs of $\calG$ assigned to the different partitions and may not be disjoint (i.e., same node may be contained in two or more partitions because of replication).
        Given a hyperedge $e$,
        this procedure finds a minimal subset of the partitions $MD_e \subseteq \calG$, such that every node in $e$ is contained 
        in at least one partition in $MD_e$. We use the greedy Set Cover algorithm
        for this purpose. We start with the partition $G_i$ that has the maximum overlap with $e$, and include it
        in $MD_e$. We then remove all the nodes in $e$ that are contained in $G_i$ (i.e., ``covered'' by $G_i$) and 
        repeat till all nodes are covered.
    \item {\bf getQuerySpan($\calG, e$):} Given a current placement $\{G_1, \cdots, G_N \}$ and a hyperedge $e$, 
        this procedure finds the span of the hyperedge $e$. We use the same algorithm as above, but return $|MD_e|$
        instead of $MD_e$.
    \item {\bf getAccessedItems($\calG, e, g \in \calG$):} Given a current placement $\calG = \{G_1, \cdots, G_N \}$, a hyperedge $e$ and
        a partition $g \in \calG$, this returns the set of items that the query corresponding to $e$ would access from partition $g$, 
        as computed by the greedy Set Cover algorithm. This may be empty even if $e \cap g \ne \phi$.
    \item {\bf pruneHypergraphBySpan($\calG, \calH, minSpan$):} Given a current placement $\calG$ and 
        a value of $minSpan$, this routine removes all hyperedges from \calH\ with span less than 
        or equal to $minSpan$.
    \item {\bf getKDensestNodes($\calH, K$):} Given a hypergraph \calH, this procedure returns a 
        dense subgraph containing at nodes having total weight of atmost $K$. We use a greedy algorithm for this purpose: 
        we find the lowest degree node and remove that node and all edges incident on it; if the 
        graph still has nodes having total weight more than $K$, we repeat the process by finding the lowest degree
        node in the new graph.
    \item {\bf pruneHypergraphToSize($\calH, K$):} Given a current placement $\calG$ and 
        a value of $K$, this routine uses the same algorithm as for getKDensestNodes to find a (dense) hypergraph
        over nodes having total weight of $K$. 
    \item {\bf totalWeight($V$, $W_v$):} Given a set of vertices $V$ and weight vector of vertices $W_v, v\in V$, this routine returns the total weight of vertices. 
\end{list}
We note that, because of the modularized way our framework is designed, we can easily use different, more efficient algorithms for
solving these subproblems.

\subsection{Iterative HPA (IHPA)}
\label{sec:multiprune}
Here, we start by using HPA to get a partitioning of the data items into exactly $N_e$ partitions (recall that $N_e$ is the 
minimum number of partitions needed to store the data items). We then prune the original hypergraph $\mathcal{H}(V, E)$ to
get a residual hypergraph $\mathcal{H}^{'}(V^{'}, E^{'})$ as follows: we remove all hyperedges that are completely contained
in a single partition (i.e., hyperedges with span 1), and we then remove all the data items that are not contained in 
any hyperedge. If the number of nodes in the $\mathcal{H}'$ is less than $(N - N_e) C$ (i.e., if the data items fit in 
the remaining empty partitions), we apply HPA to obtain a balanced partitioning of $\mathcal{H}'$ and place the partitions
on the remaining partitions. This process is repeated if there are still empty partitions.

If the number of nodes in $\mathcal{H}'$ is larger than the remaining capacity, we prune the graph further by removing
the hyperedges with the lowest span one at a time (these hyperedges are likely to see the least improvement by replication) and 
the data items that now have 0 degree, until the number of nodes in $\mathcal{H}'$ becomes sufficiently low; then we apply HPA to obtain a balanced partitioning of $\mathcal{H}'$ and place the partitions on the remaining partitions. 
If there are still empty partitions, we repeat the process by reconstructing a new residual graph. Algorithm~\ref{tab:alg1} depicts the pseudocode for this technique.
%The process is repeated from beginning (removing hyperedges with span 1 and partitioning remaining hypergraph) is repeated if there are still empty partitions.

\begin{algorithm}[htb]
\small
\caption{Iterative HPA (IHPA)}
%\label{lcer}
\begin{algorithmic}[1]
\REQUIRE $\mathcal{H}(V, E), N, C$
\STATE Run HPA to get an initial partitioning into $N_e$ partitions: $\calG = \{G_1, G_2,\dots, G_{N_e}\}$;
\STATE $edgeCost =$ avgDataItemsPerQuery$(\mathcal{H})$;
\WHILE{$edgeCost\neq 0$ \textbf{and} $|\calG| \ne N$}
\STATE $\calH^{'}(V', E') =$ pruneHypergraphBySpan$(\calG, \calH, edgeCost)$;
%\STATE $N_{cur}=\frac{|V^{'}|}{C}$;
\STATE $N_{cur}=\frac{totalWeight(V^{'},W_{v'})}{C}$;
\IF{$|\calG| + N_{cur} \le N $ \textbf{and} $|\mathcal{H^{'}}|\neq 0$}
\STATE $\calG$ = $\calG$ $\cup$ HPA$(\mathcal{H^{'}}, N_{cur})$;
\ELSIF{$|\calG| + N_{cur} > N$}
\STATE $\calG$ = $\calG$ $\cup$ HPA$(\mathcal{H^{'}}, N - |\calG|)$;
\ELSE
\STATE \textbf{decrement} $edgeCost$ \textbf{by} 1;
\ENDIF
\ENDWHILE 
\RETURN final partitions $G_1, G_2,\cdots, G_{N}$
\end{algorithmic}
\label{tab:alg1}
\end{algorithm}

\subsection{Dense Subgraph-based (DS)}
This algorithm directly follows from the discussion in the previous section. As above, we use HPA to get an initial partitioning.
We then fill the remaining $N-N_e$ partitions one at a time, by identifying a dense subgraph of the residual hypergraph. This is done
by removing the lowest degree nodes from $\mathcal{H}'$ until the number of nodes in it reaches $C$ (the partition capacity). These
data items are then placed on one of the remaining partitions, and the procedure is repeated until all partitions are utilized. Pseudocode is shown in Algorithm~\ref{tab:alg2}.
\begin{algorithm}[htb]
\small
\caption{Dense Subgraph-based (DS)}
%\label{gldnr}
\begin{algorithmic}[1]
\REQUIRE $\mathcal{H}(V, E), N, C$
\STATE Run HPA to get an initial partitioning into $N_e$ partitions: $\calG = \{G_1, G_2,\dots, G_{N_e}\}$;
\STATE $\mathcal{H^{'}} = \mathcal{H}$;
\WHILE{$|\calG|\neq N$}
\STATE $\mathcal{H^{'}} =$ pruneHypergraphBySpan$(\calG, \mathcal{H}, 1)$;
\IF{$|\calH'| = 0$}
\STATE \textbf{break};
\ENDIF
\STATE $denseNodes$ = getKDensestNodes$(\mathcal{H^{'}}, C)$;
\STATE Add a partition containing $denseNodes$ to $\calG$;
\ENDWHILE 
\RETURN final partitions $G_1, G_2,\cdots, G_{N}$
\end{algorithmic}
\label{tab:alg2}
\end{algorithm}
\subsection{Pre-Replication-based Algorithm (PRA)}
This algorithm is based on the idea of identifying small separators and replicating them. However, we do not directly adapt the recursive 
algorithm described in Section \ref{sec:definition} for two reasons. First, since we have a fixed space budget for replication, we must somehow distribute this
budget to the various stages and it is unclear how to do that effectively.
More importantly, the basic algorithm of bisecting a graph
and then recursing is not considered a good approach for achieving good partitioning~\cite{271619, Karypis98multilevelk-way}.

We instead propose the following algorithm. We start with a partitioning returned by HPA, and identify ``important'' nodes such that 
by replicating these nodes, the average query span would be reduced the most.
Then, we create a new hypergraph by replicating these nodes (until we have enough nodes to fill all the partitions), and run HPA once again to attain a final partitioning. However, neither of these steps is straightforward.

\topic{\textbf{Identifying Important Nodes}}: \eat{The goal with this step is to decide replicating which nodes is likely to
bring the most benefit}The goal is to decide which nodes will offer the most benefit if replicated. We start with a partitioning obtained using HPA, and then analyze the partitions to decide this. We
describe the intuition first. Consider a node $a$ that belongs to some partition $G_i$. Now count the number of those hyperedges
that contain $a$ but do not contain any other node in $G_i$; we denote this number by $score_a$. If this number is high, then the node is a good candidate for replication since replicating the node is likely to reduce the query spans for several queries. We use the partitioning returned by HPA to rank all the nodes in the decreasing order by this count, and then process the nodes one at a time.

%Given $G$, each partition $G_j$ is analysed to get a list of nodes, where each node/data item $d_{ij}, i\in N_D, j\in N_G$ is incident upon by set of hyperedges $E_{d_{ij}}$, that span more than one partition and that contain $d_{i}$ as the only node in partition $G_j$. This list of nodes is sorted in the order of number of incident hyperedges spanning multiple partitions. Let this list be $D_{imp}$, $d_{ij}\in D_{imp}$. \newline
\begin{algorithm}[t]
\small
\caption{Pre-replication-based Algorithm (PRA)}
%\label{pir}
\begin{algorithmic}[1]
\REQUIRE $\mathcal{H}(V, E), N, C$
\STATE Run HPA to get an initial partitioning into $N_e$ partitions: $\calG = \{G_1, G_2,\dots, G_{N_e}\}$;
\FOR{$v \in V$}
\STATE {\bf let} $v$ be contained in partition $G_v$;
\STATE {\bf compute} $score_v = | \{e \in E\ |\ e \cap G_v = \{v\} \}|$;
\ENDFOR
\STATE $H^r = H$;
\FOR{$v \in V$ in decreasing order by $score_v$}
\STATE $E_{v} = \{ e \in E\ |\ v \in e \}$;
\STATE $G_{v} = \{ $getSpanningPartitions$(\calG, e)\ |\ e \in E_{v} \}$;
\STATE $S =$ getHittingSet$(G_v)$;
\FOR{$g\in S$}
\STATE $copy_g=$ makeNewCopy$(v)$;
\FOR{$e \in E_v$ {\bf s.t.} $g \in $getSpanningPartitions$(\calG, e)$}
\STATE  $e = e - \{v\} + \{copy_g\}$; 
\ENDFOR
\ENDFOR
\ENDFOR
\STATE $\calG =$ HPA$(\calH^r, N)$;
\RETURN final partitions $G_1, \cdots, G_{N}$
\end{algorithmic}
\label{tab:alg3}
\end{algorithm}

\begin{figure}[t]
\centering
\includegraphics[width=3.5in]{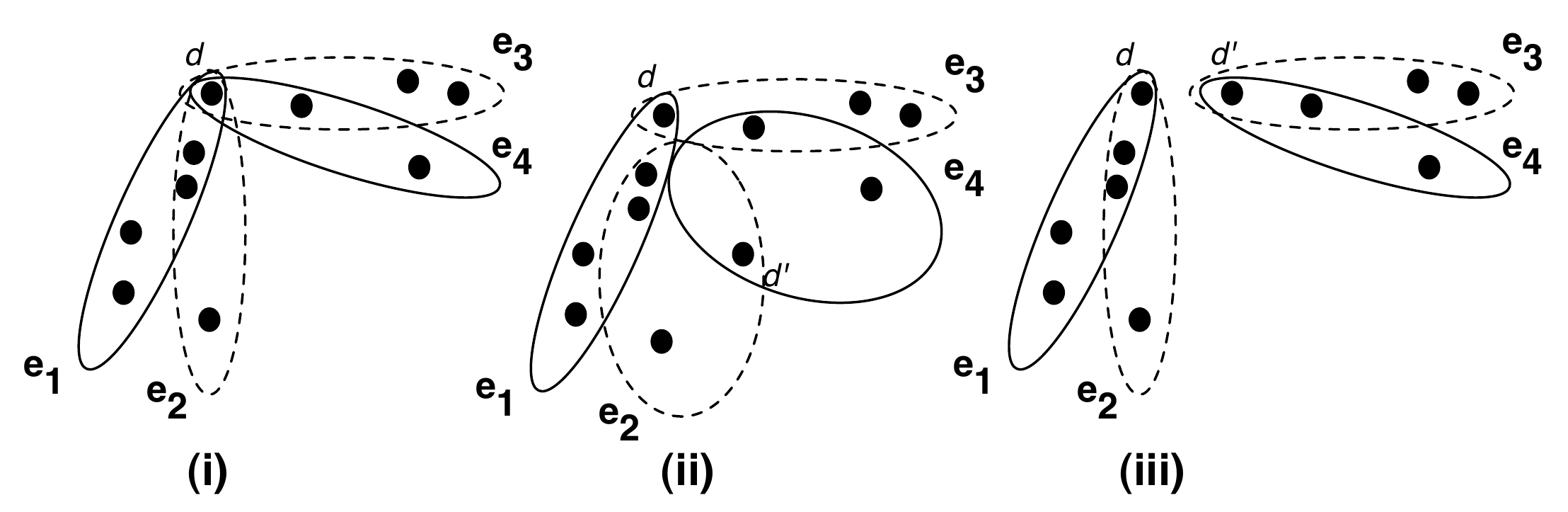}
\vspace{-5pt}
\vspace{-5pt}
\vspace{-5pt}
\caption{When replicating a node, distribution of the copies to the hyperedges must be done carefully. Distribute the replica copies such that it results in entanglement of the incident hyperedges.}
\vspace{-5pt}
\vspace{-5pt}
\label{fig:important}
\end{figure}

\topic{\textbf{Replicating Important Nodes}}: Let $d$ be the node with the highest value of $score_d$ among all nodes. We now 
have to decide how many copies of $d$ to create, and more importantly, which copies to assign to which hyperedge. Figure \ref{fig:important}(ii)
illustrates the problems with an arbitrary assignment. Here we replicate the node $d$ to get one more copy $d'$, and then we
assign these two copies to the hyperedges $e_1, e_2, e_3, e_4$ as shown (i.e., we modify some of the hyperedges to remove $d$ and add $d'$ instead).
However, the assignment shown is not a good one for a somewhat subtle reason. Since $e_1$ and $e_3$ (which are assigned the original
$d$) do not share any other nodes, it is likely that they will span different sets of partitions, and one of them is likely to 
still pay a penalty for node $d$. On the other hand, the assignment shown in Figure \ref{fig:important}(iii) is better because here
the copies are assigned in a way that would reduce the average query span.

We formalize this intuition in the following algorithm. For node $d$, let $E_d = \{e_{d_1}, e_{d_2}, \cdots, e_{d_k}\}$ denote the set of hyperedges that contain
$d$. For hyperedge $e_{d_i}$, let $\mathcal{G}_{d_i}$ denote the set of partitions that $e_{d_i}$ spans. 
We then identify a set of partitions, $S$, such that each of $\mathcal{G}_{d_i}$  contains at least one partition from this set (i.e., $S \cap \mathcal{G}_{d_i} \ne \phi$).
Such a set is called a ``hitting set''. We then replicate $d$ to make a  total of $|S|$ copies. Finally, we assign the copies to the hyperedges according to the 
hitting set, i.e., we uniquely associate the copies of $d$ with the members of $S$, and for a hyperedge $e_{d_i}$, we assign it a copy such that the associated element
from $S$ is contained in $\mathcal{G}_{d_i}$ (if there are multiple such elements, we choose one arbitrarily).

%We then identify a minimal ``hitting set'' of the collection of sets
%$\mathcal{G}_{d_1}, \cdots, \mathcal{G}_{d_k}$; this is a set of partitions such that each of $\mathcal{G}_{d_i}$ contains at least one partition from this set. We then replicate $d$
%to make as many copies as the size of the hitting set. Finally, the copies are assigned to the hyperedges according to the hitting set.

The problem of finding the smallest hitting set is NP-Hard. We use a simple greedy heuristic. We find the partition that is common to the maximum number of sets $\mathcal{G}_{d_i}$, 
include it in the hitting set, remove all sets that contain it, and repeat. Algorithm~\ref{tab:alg3} depicts the pseudocode for this technique.

\subsection{Local Move Based Replication (LMBR)}
\label{sec:lmbr}
\begin{figure}[t]
\centering
\vspace{-5pt}
\includegraphics[width=3.4in]{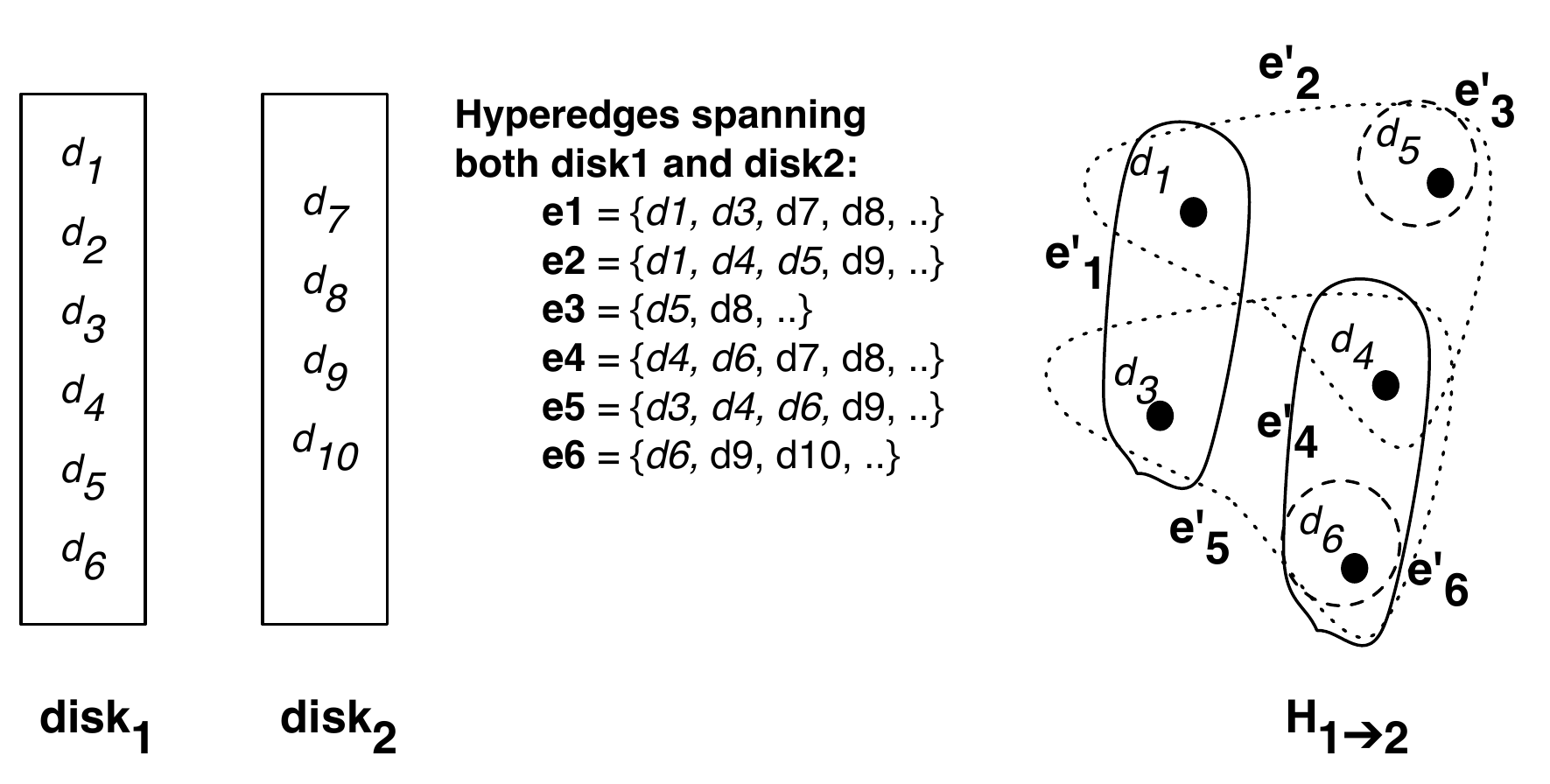}
\vspace{-5pt}
\vspace{-5pt}
\vspace{-5pt}
\caption{Constructing $H_{1 \rightarrow 2}$: e.g., corresponding to hyperedge $e_1$
that spans both partitions, we have a hyperedge $e'_1$ over $d_1$ and $d_3$.}
\vspace{-5pt}
\vspace{-5pt}
\label{fig:h12}
\end{figure}

Finally, we consider algorithms based on local greedy decisions about what to replicate, starting with a partitioning
returned by HPA. For simplicity and efficiency, we chose to employ moves involving two partitions. More specifically, at each step, 
we copy a small group of data items from one partition to another. The decisions are made greedily by finding the move that results
in the highest decrease in the average query span (``benefit'') per data item copied (``cost'').
For this purpose, 
at all times, we maintain a priority queue containing the best moves from $partition_i$ to $partition_j$, for all $i \ne j$. 
For two partitions $partition_i, partition_j$, the best group of data items to be copied from $partition_i$ to $partition_j$ is calculated
as follows. Let $E_{ij} = \{e_{{ij}_1}, \cdots, e_{{ij}_l}\}$ denote the hyperedges that contain data items from both
the partitions. We construct a hypergraph $H_{i \rightarrow j}$ on the data items of $partition_i$ as follows: for every edge $e_{{ij}_k}$, we add a 
hyperedge to $H_{i \rightarrow j}$ on the data items common to $e_{{ij}_k}$ and $partition_i$. Figure \ref{fig:h12} illustrates this process with
an example.

Now, if we were to copy a group of data items $X$ from $partition_i$ to $partition_j$, the resulting decrease in total span (across
all edges) is exactly the number of hyperedges in $H_{i \rightarrow j}$ that are completely contained in $X$. Thus, the problem of finding
the best move from $partition_i$ to $partition_j$ is similar to the problem of finding a dense subgraph, with the main difference
being that, we want to minimize the cost/benefit ratio and not maximize the benefit alone. 
Hence, we modify the algorithm for finding dense subgraph as follows. We first compute the cost/benefit ratio for the entire group
of nodes in $H_{i \rightarrow j}$. The cost is set to $\infty$ if the number of nodes to be copied is more than the empty space in $partition_j$.
We then remove the lowest degree node from $H_{i \rightarrow j}$ (and any incident hyperedges), and again compute the cost/benefit ratio. 
We pick the group of items that results in the lowest cost/benefit ratio.

After finding the best moves for every pair of partitions, we choose the overall best move, and copy the data items accordingly.
We then recompute the best moves for those pairs which were affected by this move (i.e., the pairs containing the destination
partition), and recurse until all the partitions are full.

\topic{Improved LMBR:} Although the above looks like a reasonable algorithm, it did not perform very well in our first
set of experiments. As described above, the algorithm has a serious flaw. Going back to the example in Figure \ref{fig:h12},
say we chose to copy data item $d_6$ from $partition_1$ to $partition_2$. In the next step, the same move would still 
rank the highest. This is because the construction of hypergraph $H_{1 \rightarrow 2}$ is oblivious to the fact that $d_6$
is also now present in $partition_2$. Further, it is also possible that, because of replication, neither of the partitions is
actually accessed at all when executing the queries corresponding to $e_4, e_5$ or $e_6$.

To handle these two issues, during the execution of the algorithm, we maintain the exact list of partitions that would be activated 
for each query; this is calculated using the Set Cover algorithm described in Section \ref{sec:definition}. Now when we consider whether to
copy a group of items from $partition_i$ to $partition_j$, we make sure that the benefit reflects the actual query span reduction given
this mapping of queries to partitions. Pseudocodes for this algorithm is give in Algorithm~\ref{tab:alg4} and \ref{tab:alg5}.

\begin{algorithm}[t]
\small
\caption{Improved LMBR}
%\label{lgrasc}
\begin{algorithmic}[1]
\REQUIRE $\mathcal{H}(V, E), N, C$
\STATE Run HPA to get initial partitions $\calG = \{G_1, G_2,\dots, G_{N}\}$ into $N$ partitions;
\STATE Compute the set cover $MD_e$ for each query $e$;
\STATE Initialize PQ (priority queue) to empty;
\FOR{$g = G_1$ to $G_N $}
\FOR{$g' = G_1$ to $G_N, g \neq g'$}
%\STATE $\forall e \in E_{md}\subset E, \{partition_1, partition_2\}\in MD_e$ \textbf{do}
\STATE PQ.insert($g\rightarrow g'_,$ maxGain$(\calG, g,  g')$);
\ENDFOR
\ENDFOR
\WHILE{\emph{all partitions are not full}}
\STATE ($g_{src} \rightarrow g_{dest}$) = PQ.bestMove();
\STATE \textbf{copy} appropriate items from $g_{src}$ to $g_{dest}$;
%\STATE \textbf{update} $MD_e$; %, E_{md}$;
\FOR{$g=G_1$ to $G_N$, $g \ne g_{dest}$}
%\STATE $\forall e \in E_{md}\subset E, partition_1\in MD_e$ \textbf{do}
\STATE PQ.update($g \rightarrow g_{dest}$, maxGain(\calG, $g,  g_{dest}$));
\STATE PQ.update($g_{dest}\rightarrow g$, maxGain(\calG, $g_{dest},  g$));
\ENDFOR
\ENDWHILE
\RETURN final partitions $G_1, \cdots, G_N$; %$G^{'}$
\end{algorithmic}
\label{tab:alg4}
\end{algorithm}
\setlength{\textfloatsep}{10pt}
\begin{algorithm}[t]
\small
\caption{Improved LMBR maxGain Method}
%\label{gain}
\begin{algorithmic}[1]
    \REQUIRE $\calG=\{G_1, \cdots, G_{N}\}, \calH(V, E), G_{src} \in \calG, G_{dest} \in \calG$
    \STATE $E_{src} = \{e \in E\  |$  getAccessedItems$(\calG, e, G_{src}) \ne \phi\}$;
    \STATE $E_{dest} = \{e \in E\  |$  getAccessedItems$(\calG, e, G_{dest}) \ne \phi\}$;
\STATE $E = E_{src}\cap E_{dest}$;
\IF{$|E|\neq 0$}
\STATE $V' = \cup_{e \in E}\  $getAccessedItems$(\calG, e, G_{src})$;
\STATE $E' = \{ $getAccessedItems$(\calG, e, G_{src})  | e \in E \}$;
\STATE {\bf create hypergraph} $\calH'(V', E')$;
\STATE $C_{dest} = C - |G_{dest}|$;
\IF{$C_{dest}\neq 0$}
\STATE $H'=$ pruneHypergraphToSize$(H', C_{dest})$;
\WHILE{$|H'| > 0$}
\STATE {\bf compute} gain = $|E'| / |V'|$
\STATE {\bf remove} lowest degree node from $H'$ and incident edges;
\ENDWHILE 
\ENDIF
\ENDIF
\RETURN the best value of gain found in the process and the corresponding $V'$;
\end{algorithmic}
\label{tab:alg5}
\end{algorithm}
\subsection{3-Way Replication Algorithms}
\label{sec:3-way}
As we have already discussed, many large-scale data management systems provide default 3-way replication. 
%If this default replication is done smartly then we can reap additional benefits such as minimizing distributed overheads and energy consumption. 
Here we briefly discuss how the algorithms described above can be modified to handle 3-way replication.

\topic{PRA-Based 3-Way Replication: } We identify PRA the most suitable algorithm to do this effectively, and modify PRA as follows. %We slightly modify PRA algorithm to support 3-way replication. 
Because we are interested in replicating all the nodes 3-way, we eliminate the step of finding important nodes from PRA and we replicate each node 3-way by using our ``hitting set`` technique to decide which copy must be shared with what hyperedges. PRA basically
aims to separate the incident hyperedges in the hypergraph by distributing the copies of node $d$ smartly to incident hyperedges. 
%Every time PRA disentangles $|E_d|$ incident hyperedges, span of each incident hyperedge $e\in E_d$ reduces by one.

\topic{Simple Distribution Algorithm: }In this algorithm, for each node $d$ in the hypergraph we find the set of incident hyperedges $E_d$. We assign 3 copies of $d$ among $|E_d|$ edges randomly, by assigning every $\frac{|E_d|}{3}$ hyperedges single copy of $d$.
Only difference between this algorithm and PRA based 3-way replication algorithm is that PRA based algorithm makes best effort to distribute the copies of node $d$ among incident hyperedges $E_d$.% such that they are disentangled in-graph; whereas this algorithm doesn't provide any guarantees on the disentanglement of incident hyperedges. Disentanglement of incident hyperedges is important, as it improves the chances of HPA to find the minimize cuts because HPA uses heuristics to find the min-cuts.

\topic{IHPA-Based Algorithm: } In IHPA for 3-way replication we run HPA to get partitioning without replication. We remove all the hyperedges with span 1 from the input graph, and 
run HPA again on the residual graph to get additional partitions. We repeat this process one
more time to replicate each node exactly 3 times. %Repeat IHPA multiple times and at each step remove the nodes that have been replicated thrice from the input graph and the hyperedges spanning one partition. 
%It is possible that IHPA will do partitioning on all the given partitions for 3-way replication without replicating all the nodes 3-way. So some of the partitions may have free space. So, for better bin packing DS based technique can be used in conjunction with IHPA. Remove already 3-way replicated nodes from the input graph and given some additional space $K$ in a partition, use DS based technique to find $K$ most dense graph and place it into the partition.

\subsection{Discussion}
We presented four heuristics for data placement with replication. There are clearly many other variations of these 
algorithms, some of which may work better for some inputs, that can be implemented quickly and efficiently using 
our framework and the core operations that it supports (e.g., finding dense subgraphs). In practice, taking the best
of the solutions produced by running several of these algorithms would guarantee good data placements.

Finally, while describing the algorithms, we assumed a homogeneous setup where all partitions are identical and all 
data items have equal size. We have also extended the algorithms to the case of heterogeneous data
items. The hMETIS package that we use and also other hypergraph partitioning packages, allow the nodes to have weights. 
For heterogeneous case the dense subgraph algorithm is modified to account for the weights, by removing the node with the lowest 
value of degree till we have nodes having total specified weight (for both DS and LMBR). Similarly, PRA is modified by allowing the replication in the original hypergraph such that total weight of replicated nodes is no greater than the sum of all extra available partition capacities. We omit the full details due to lack of space.

\section{Experimental Evaluation}
\label{sec:expts}

\begin{figure*}[t]
\vspace{-10pt}
\centering
\subfigure[]{
\includegraphics[scale=0.25]{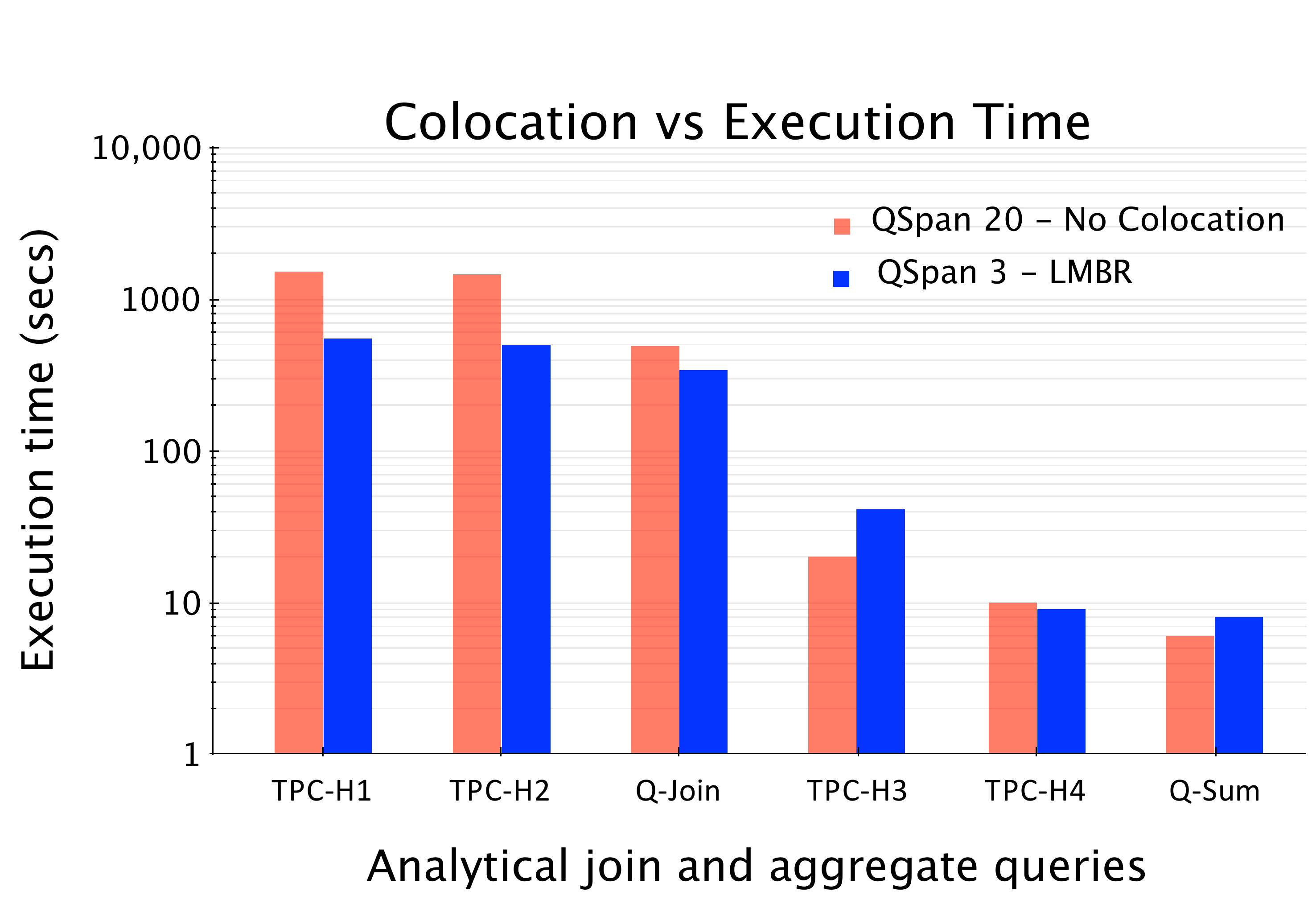}
\label{fig:execution_tpch}
}
\hspace{0.6in}
%\hspace{.1in}
%\subfigure[]{
%\includegraphics[scale=1.10, trim=0 0 10 0]{het2.pdf}
%\label{fig:het2}
%}
%\hspace{.1in}
\subfigure[]{
\hspace{-.3in}
\includegraphics[scale=0.25]{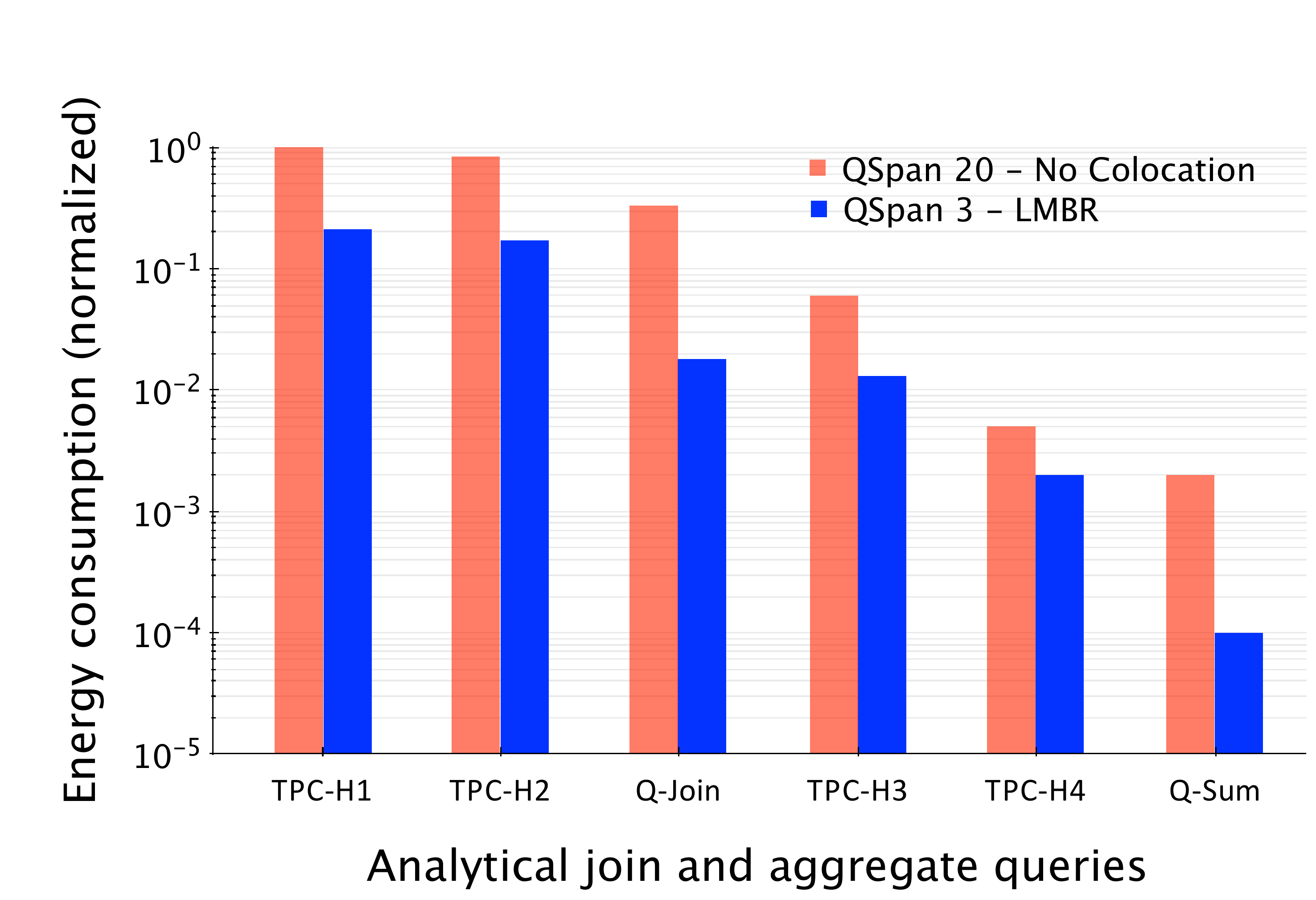}
\label{fig:energy_tpch}
}
\vspace{-10pt}
\caption[]{Experiments on a TPC-H Benchmark showing effect of co-location on query response times and resource consumption.} % Results shows that irrespective of type of query, energy consumption decreases significantly with co-location of accessed data items. }
\label{fig:span_res}
\vspace{-15pt}
\end{figure*}

We begin with presenting the result of a set of experiments designed to evaluate the effects of
query span on query response times and resource consumption, to further bolster our claim that
minimizing query span typically leads to reduced resource consumptions. We then present an extensive
set of experiments evaluating the effectiveness of our algorithms at minimizing the query span for a
collection of synthetic and real workloads.

\subsection{Query Span and Resource Consumption}
%We evaluate the effects of co-location in terms of query response time and resource consumption. 
We conducted a set of experiments analyzing the effect of query span on the total amount of resources consumed, and the total energy consumed, under a variety of settings. We performed this experiment on 20 Amazon EC2 medium instances. 
We use the same two settings and the same set of queries that we used in the experiments presented
in Section \ref{sec:intro-expts}. 
The first setting is a horizontally
partitioned MySQL cluster, where we evaluate two complex analytical join queries (TPC-H1, TPC-H2),
and two single-table aggregate queries (TPC-H3, TPC-H4), on a TPC-H dataset. 
%Two of the queries are complex analytical join queries (TPC-H1, TPC-H2), whereas the other two are simple aggregation queries (TPC-H3, TPC-H4) on a single table. 
The second setting is a homegrown distributed query processor that sits atop
multiple MySQL instances running on a cluster where predicate evaluations are pushed on to the individual nodes and data is shipped to 
a single node for perform the final steps.  %master then performs the final processing. 
We evaluate a complex join query (Q-Join) and a single-table aggregate query (Q-Sum) on that setup.
%On this setup we evaluate two queries: a complex join query (Q-Join)
%and a simple aggregate query on a single table (Q-Sum). 

To compare the cost of our best co-location scheme LMBR, we run around 10000 additional queries with
TPC-H1, TPC-H2, TPC-H3, TPC-H4 , Q-join and Q-Sum on our setup, so that we can construct the
hypergraph of these queries. We then perform  min-cut partitioning over this hypergraph to get a
20-way partitioning, and then we apply LMBR on this setup. Based on placement given by LMBR, we
place the data items across the 20 machines. Then we execute our test queries and carefully make 
sure that each query is executed on the set of machines that it spans. Query span is calculated 
by using set-cover algorithm on the placement suggested by LMBR. Average span over these test
queries was 3, i.e., data needed for these queries were located on an average of 3 machines 
using LMBR. 

In Figure~\ref{fig:execution_tpch}, we plot the query response times of our test queries on the
horizontal partitioning placement on 20 machines and we compare it with the query response times
when executed on LMBR-suggested placement. We notice that query response times for complex
analytical test queries TPC-H1, TPC-H2 and Q-join decrease significantly when executed on LMBR
suggested placement. This is because of minimization of overheads caused by distributed analytical
processing, e.g., communication overheads in processing complex joins. On the other hand, query response times for test queries TPC-H3, TPC-H4 and Q-Sum increase with co-location. 
This confirms our intuition that parallelism is more effective for simple queries than for complex
queries.
%This may be because of reduction in amount of parallelism when span is reduced to 3 machines. 

Figure~\ref{fig:energy_tpch} shows that, irrespective of the type of the query, energy consumption
decreases significantly with co-location of accessed data items. It shows that most reduction in
energy consumption for complex analytical query is for TPC-H1that is almost $79\%$, whereas for
Q-Join we observe $31\%$ reduction. For simple aggregate queries, firstly we observe that there can
be a tradeoff between query response time and energy consumption on co-location. Secondly, for
queries TPC-H3, TPC-H4 where reduction in energy consumption is $77\%$ and $57\%$ and for Q-Sum we
observe $71\%$. Depending upon the optimization goal such as query response time or energy
minimization or both, one may choose to colocate the data items or not. In this work, we
specifically focus at opportunities where co-location is applicable and provides us significant
benefits in terms of minimization of energy consumed per query, it may also minimize query response
times, for example: in case of complex analytical queries.

This experiment highlights the fact that, query response time may increase or decrease with
co-location depending up on the nature of the query (complex analytical or simple aggregate). But in
all cases, energy costs reduces with a good data co-location, for example: co-location provided by
LMBR.

\eat{In Figures~\ref{fig:mot1} and \ref{fig:mot2}, we plot the execution times and the energy consumed as the number of machines across
which the tables are partitioned (and hence query span) increases. As we can see, the execution times of the TPC-H
queries run on MySQL cluster actually increased with parallelism, which may be because of nested loop join implementation in MySQL cluster (a known problem that is being fixed). 
Our implementation shows that execution time remains constant, but in all cases, energy costs increase with query span.
In the second experiment with simpler queries (Figures~\ref{fig:mot3} and \ref{fig:mot4}), though execution times decrease as the query span increases, energy consumption increases in all cases. 
The energy consumed is computed using the Itanium server power model
calculated by using Mantis full-system power modelling technique~\cite{Economou06full-systempower}. 
We use the \emph{dstat} tool to collect various system performance counters such as CPU utilization, network read and writes, I/O, and memory footprint, which along
with the power model is used to compute the total energy consumed. }

\subsection{Query Span Experiments}
We evaluate effectiveness of our proposed algorithms by building a trace-driven simulator to experiment with different data placement 
%and scheduling
policies. The simulator instantiates a number of partitions as needed by the experimental setup, uses
a data placement algorithm for distributing the data among the partitions, and 
replays a query trace against it to measure the query span profiles. \eat{For each partition, we use
the formulas and the parameters discussed in Section \ref{epq} to calculate the energy
consumption.}

We conducted an extensive experimental study to evaluate our algorithms, using several real and
synthetic datasets. Specifically, we used the following three datasets:
\begin{list}{$\bullet$}{\leftmargin 0.15in \topsep 1pt \itemsep 0pt}
    \item \underline{Random:} Instead of generating a query workload completely randomly, we use a different
        approach to better understand the structure of the problem. We first generate a random
        {\em data item graph} of a specified density (edges to nodes ratio). We then randomly
        generate queries such that the data items in the query form a connected subgraph in 
        the data item graph. For low density data item graphs, this induces significant structure
        in the query workload that good data placement algorithms can exploit for better performance. \eat{Figure \ref{fig:snowflake}(a) shows an example data object graph where the numbers indicate the data item sizes (in MB). Figure \ref{fig:snowflake}(b) shows several queries that may be generated using this data item graph -- each of the queries forms a connected subgraph in the data item graph.}
    \item \underline{Snowflake:} This is a special case of the above where the data item graph is a tree. This
        workload attempts to mimic a standard SQL query workload.\eat{An example data item graph corresponding to the Snowflake dataset is shown in Figure \ref{fig:snowflake}(c).} 
        %Here the large squares indicate the first-level relations, and the small squares indicate the second-level relations. 
        We treat each column of each relation as a separate data item. An SQL query over such a schema that does not contain a Cartesian product corresponds to a connected subgraph in this graph.
    \item \underline{ISPD98 Benchmark Data Sets}: In addition to the above synthetic datasets, 
        we tested our algorithms on standard ISPD98 benchmarks~\cite{274546}. ISPD98 circuit benchmark suite contains 18 
        circuits ranging from 12,752 to about 210,000 nodes. Hypergraph density (hyperedges to nodes ratio) in all the ISPD98 circuit 
        benchmarks is close to 1, i.e., these graphs are quite sparse. We show results for the first 10 circuit datasets, that contain 12,752 to 69,429 nodes.
\end{list}
%Due to space constraints, we present only the results for the Random datasets here. Experiments
%for the other two datasets can be found in the appendix.
%
\vspace{10pt}
We compare the performance of six algorithms: (1) {\bf Random}, where the data is replicated and
distributed randomly, (2) {\bf HPA}, the baseline hypergraph partitioning algorithm, (3-6) the four
algorithms that we propose, {\bf IHPA, PRA, DS,} and {\bf LMBR} (Section \ref{sec:algorithms}).
We use the hMETIS hypergraph partitioning algorithm~\cite{Karypis99multilevelhypergraph,hMETIS} 
as our HPA algorithm. 
The experiments were run on a Intel Core2 Duo CPU 2.10GHz, 4GB RAM, Windows PC running Windows 7.
All plotted numbers (except the numbers for the ISPD98 benchmark) are averages over 10 random
runs. For reproducibility, we list the values of the remaining hMETIS parameters: {\em Nruns} = 20, {\em
CType} = 2, {\em RType} = 1, {\em VCycle} = 1, {\em Reconst} = 1, {\em dbglvl} = 0.

The key parameters of the dataset that we vary are: (1) {\em $|$D$|$}, the number of data items,
(2-3) {\em minQuerySize} and {\em maxQuerySize}, the bounds on the query sizes
that are generated, (4) {\em NQ}, the number of queries, (5) {\em C}, the partition capacity, 
(6) {\em numPartitions (NPar)}, the number of partitions, and 
(7) {\em density} of the data item graph (defined to be the ratio of the number of edges to the 
number of nodes). The default values were: 
$|$D$|$ = 1000, 
minQuerySize = 3, maxQuerySize = 11,
NQ = 4000,
C = 50, NPar = 40, and density = 20.

%{\small
%\begin{table}[!]
%\small
%\centering
%\tabcolsep 5.8pt
%\begin{tabular}{|l|p{5cm}|l|} 
%\hline 
%\multicolumn{2}{|c|}{\textbf{hMETIS ($HPA$) Parameter Values}} \\ 
%\hline \textbf{Parameters} & \textbf{value}\\ 
%\hline\hline {$noPartitions$} 
%& Varies\\ 
%\hline {$UBfactor$} 
%& 1 for almost balanced partitioning, else varies\\ 
%\hline {$Nruns$}
%& 50\\ 
%\hline {$CType$}
%& 2\\ 
%\hline {$RType$}
%& 1\\
%\hline {$VCycle$}
%& 1\\
%\hline {$Reconst$}
%& 1\\
%\hline {$dbglvl$}
%& 0\\
%\hline 
%\end{tabular}
%\caption{$HPA$ Parameter Values}
%\label{table:hMETISpar}
%\end{table}
%}

\begin{figure*}[!]
\centering
\subfigure[]{
\includegraphics[scale=0.85]{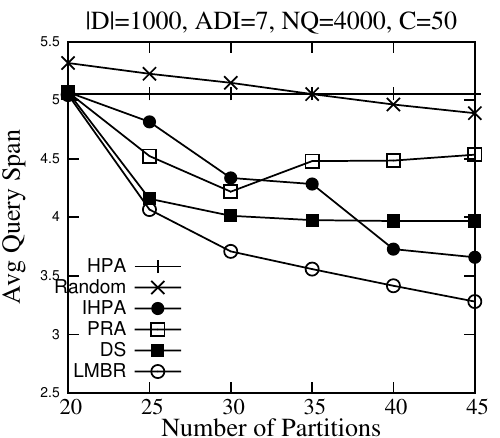}
\label{fig:exp1}
}
% \hspace{.01in}
% \subfigure[]{
% \includegraphics[scale=1.15, trim=0 0 10 0]{exp11.pdf}
% \label{fig:exp11}
% }
\hspace{-.12in}
\subfigure[]{
\includegraphics[scale=0.85]{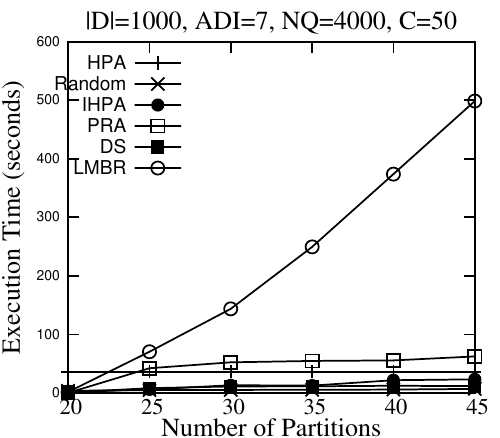}
\label{fig:exp12}
}
\hspace{-.12in}
\subfigure[]{
\includegraphics[scale=0.85]{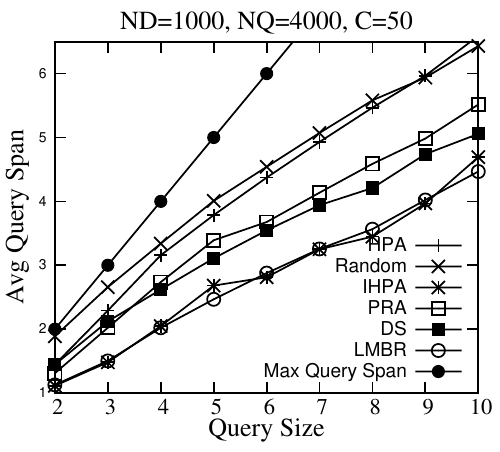}
\label{fig:exp2}
}
\hspace{-.12in}
\subfigure[]{
\includegraphics[scale=0.85]{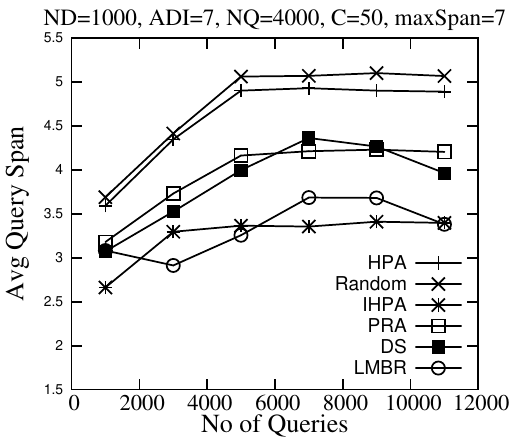}
\label{fig:exp21}
}
\vspace{-.12in}
\\
\subfigure[]{
\includegraphics[scale=0.82]{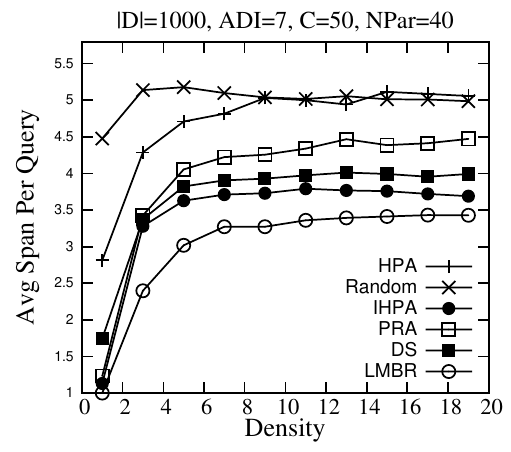}
\label{fig:density}
}
\hspace{-.12in}
\subfigure[]{
\includegraphics[scale=0.82]{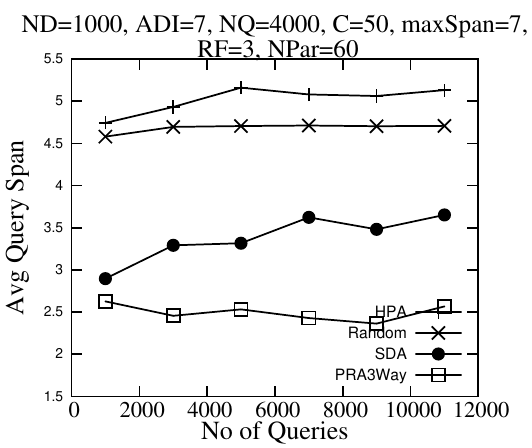}
\label{fig:noquery3way}
}
\hspace{-.12in}
\subfigure[]{
\includegraphics[scale=0.82]{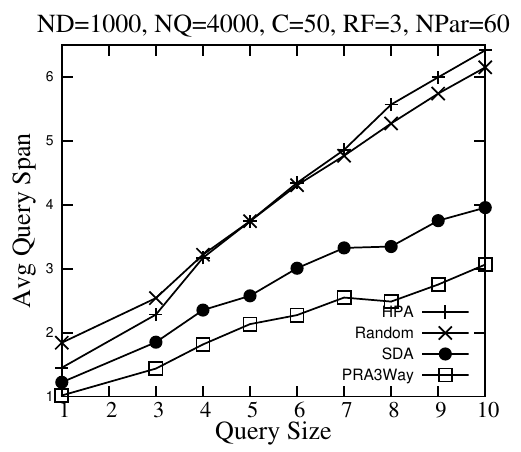}
\label{fig:lenquery3way}
}
\hspace{-.12in}
\subfigure[]{
\includegraphics[scale=0.85]{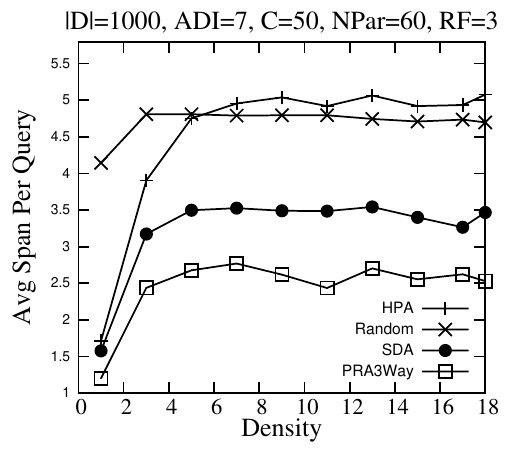}
\label{fig:density3way}
}

\vspace{-.15in}
\caption[]{$(a)-(e)$ Results of the experiments on the Random dataset with homogeneous data items illustrate the benefits of intelligent data placement with replication; 
the LMBR algorithm produces the best data placement in almost all scenarios. Note that, for clarity, the $y$-axes for several of the
graphs do not start at 0. $(f)-(h)$ 3-way replication results with replication factor of each node $RF=3$.}
\label{fig:experiments}
\vspace{-.15in}
\end{figure*}

In several of the plots, we also show the average number of data items per query, denoted {\em ADI}. 

%We conduct several experiments to evaluate our algorithms. We conduct following experiments by
%varying various parameters and draw several insights regarding the behaviour of the algorithms with
%their energy efficiency capabilities and execution time. Our algorithms take the partitions $G$
%returned by the hypergraph partitioning algorithm HPA as input and given the extra space, these
%algorithms replicate the nodes to improve the partition quality and minimize an average query span.
%Basically, HPA acts as the black box to the algorithms. 
%Since HPA can be any hypergraph partitioning
%algorithm, we use standard multilevel hypergraph partition algorithm
%\cite{Karypis99multilevelhypergraph} provided by hMETIS (A Hypergraph Partitioning Package)
%\cite{} that provides various hypergraph partitioning algorithms. Partitioning parameters
%\cite{hMETISmanual} and their values of $HPA$ used throughout the experiments is shown in Table
%\ref{table:hMETISpar}. All the experiments conducted are the average taken over 10 random datasets
%generated using the technique discussed in Section \ref{sec:simulation}.

\subsubsection{Random Dataset}
\label{sec:random}
We begin with showing the results for the Random dataset with homogeneous data items.

\topic{\textbf{Increasing Number of Partitions ($ND$)}}: First, we run experiments with increasing
the number of partitions. With the default parameters, a minimum of 20 partitions are needed to store the
data items. We increase the number of partitions from 20 to 45, and compute the average query spans,
\eat{average energy consumption,} and average execution times, for the six algorithms over 10 runs. 
%In this experiment, we choose minimum of 20 partitions ($minPartitions$), that is required to
%place the data items at-most once whereas maximum number of partitions ($maxPartitions$) as 50. HPA places
%partitions on 20 partitions, and for other algorithms we add extra partitions from 20 to 50 by adding 5 partitions
%at a time. 
Figures \ref{fig:exp1}, and \ref{fig:exp12} show the results of the experiment. % with increasing number of extra partitions. We fix the number of partitions for HPA, that is 20 that's the reason HPA displays a horizontal
%straight line. 
HPA does not do replication, and hence the corresponding plot is a straight line. The performance of
the rest of the algorithms, including Random, improves as we allow for replication. Among those, 
LMBR performs the best, with IHPA a close second. We saw this behavior consistently across almost
all of our experiments (including the other datasets). LMBR's performance does come with a
significantly higher execution times as shown in Figure \ref{fig:exp12}. This is because LMBR
tends to do a lot of small moves, whereas the other algorithms tend to have a small number of
steps (e.g., DS runs the densest subgraph algorithm a fixed number of times, whereas PRA only has 
three phases). Since data placement is a one-time offline operation, the high 
execution time of LMBR may be inconsequential compared to the reduction in {\em query span} it guarantees.

%Now we try to add extra partitions for other algorithms and try to study their behaviour
%and performance. We find that, all the replication algorithms work better than HPA, when extra space
%is available. LMBR shows the best performance and improves the quality of the partitions quite
%rapidly. Reason is that LMBR guarantees the reduction in span at every replication move. We vary the
%\emph{capping} parameter or lower bound on the number of nodes that can be replicated at a time for
%LMBR to improve it's execution time. Figure \ref{fig:exp1} and Figure \ref{fig:exp12} show that
%capping definitely reduces the execution time by far but at the cost of the quality of the solution.
%LMBR with 0 capping gives the best solution than 5 and 15 capping ones, followed by DS, IHPA, PIR
%(in that order). Figure \ref{fig:exp11} shows that LMBR is the most energy efficient replication
%algorithm whereas it is quite slow in execution time compared to other replication algorithms, but
%since these algorithms are meant to be run offline slow processing time should not be a bottleneck.

\topic{\textbf{Increasing Query Size ($ADI$)}}: Second, we vary the number of data
items per query from 2 to 10 (by setting minQuerySize = maxQuerySize), choosing the default values
for the other parameters. 
%Constant parameters are $ND=1000$, $NQ=4000$, $totalPartitions=40$, HPA runs on 20 partitions. 
As expected (Figure \ref{fig:exp2}), the average span \eat{and hence the average energy per query, both} increase
rapidly as the query size increases. The relative performance of the different algorithms is
largely unchanged, with LMBR and IHPA performing the best.
%Figure \ref{fig:exp2}
%shows the increase in query partition span with the increase in the number of data items accessed by the
%query. Provided, double the number of extra space or partitions our best replication algorithm (LMBR)
%provides 30-50\% energy savings. Other algorithms can also provide very good energy savings
%depending on the nature of the workload, extra storage and other parameters. For example: at 40
%partitions, IHPA works as good as LMBR as shown in Figure \ref{fig:exp2}.

\topic{\textbf{Increasing Number of Queries ($NQ$)}}: Next, we vary the number of
queries from 1,000 to 11,000, thus increasing the density of the hypergraph (Figure \ref{fig:exp21}). The average
query span increases rapidly in the beginning and much more slowly beyond 5,000 queries. 
Once again the LMBR algorithm finds the best solution by a significant margin compared
to the other algorithms.

\topic{\textbf{Increasing Data Item Graph Density}}: Finally, we vary the data item graph density
while from 2 (very sparse) to 20 (dense). The number of partitions was set to 40. As we can see 
in Figure \ref{fig:density}, 
for low density graphs, the average \eat{energy consumption} span of the queries is quite low, and it 
increases rapidly as the density increases. Note that the average query size did not change, 
so the performance gap is entirely because of the structure of the query hypergraph for low
density data item graphs. Further, we note that the curves flatten out as the density increases,
and don't change significantly beyond 10, indicating that the query workload essentially 
looks random to the algorithms beyond that point.

Overall, our experimental study indicates that LMBR, despite its high running time, should be the
data placement algorithm used for minimizing query span/multi-site overheads and energy consumption
in such scenarios (where we do not have any constraints on the number of replicas that must or 
can be created).

\subsubsection{3-Way Replication}
Figures~\ref{fig:noquery3way}, \ref{fig:lenquery3way} and \ref{fig:density3way} show a set of experimental results comparing the 3-way replication algorithms that we have discussed in Section~\ref{sec:3-way}.

\topic{\textbf{Increasing Number of Queries ($NQ$)}}: Increasing the number of queries, thus increasing the density of the graph, we observe that PRA based 3-way replication algorithm performs the best. This is in comparison with HPA (no replication), Random 3-way replication and simple distribution algorithm (SDA). As the number of hyperedges increases in the graph average number of hyperedges incident per node also increases. This effects the SDA algorithm, because SDA tries to distribute the 3 copies of the node randomly to the number of hyperedges incident on it. So as average number of incident hyperedges per node increases, it is more likely for SDA to make bad decisions about distribution of replicas among incident hyperedges, hence SDA's average span increases with number of queries. On the other hand, PRA employs \emph{hitting set} technique to do a more smarter replica distribution among the incident hyperedges. Increase in number of queries doesn't seem to effect the query span for PRA, which indicates the effectiveness of PRA approach. Hence, PRA based technique performs consistently better than SDA in this experiment.

\topic{\textbf{Increasing Query Size ($ADI$)}}: Query span for all the algorithms increases with an increase in average data items per query. As we saw that density of the hypergraph affects PRA and SDA, where increase in density doesn't affect PRA. \eat{basically aim to entangle the incident hyperedges in the hypergraph by distributing the copies of node $d$ to incident hyperedges. Every time PRA and SDA entangles $|E_d|$ incident hyperedges, span of each hyperedge $e\in E_d$ reduces by one.}In this experiment increase in hyperedge size doesn't affect the density of the hypergraph. Hence query span increases for SDA and PRA. PRA again performs consistently better than other algorithms.

\topic{\textbf{Increasing Data Item Graph Density}}: PRA again performs the best compared to Random and SDA when density of the graph is varied. Analysis is similar to what we have discussed before in Section~\ref{sec:random}.

We do not compare with LMBR for this scenario due to its high running time, and because it
cannot guarantee the replication constraint of 3-way replication. 

\subsubsection{Snowflake Dataset}
\begin{figure}[t]
\centering
\subfigure[]{
\hspace{-.1in}
\includegraphics[scale=0.94]{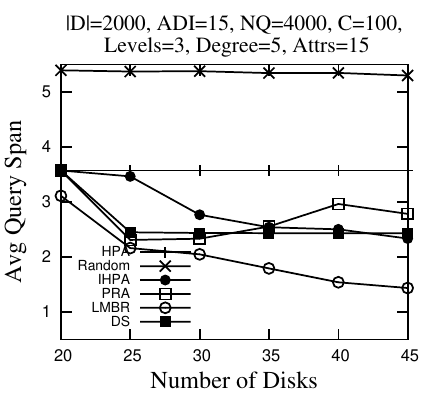}
\label{fig:snow1}
}
\hspace{-.1in}
%\hspace{.01in}
%\subfigure[]{
%\includegraphics[trim=0 0 10 0]{snow2.pdf}
%\label{fig:snow2}
%}
%\hspace{.01in}
\subfigure[]{
\includegraphics[scale=0.94]{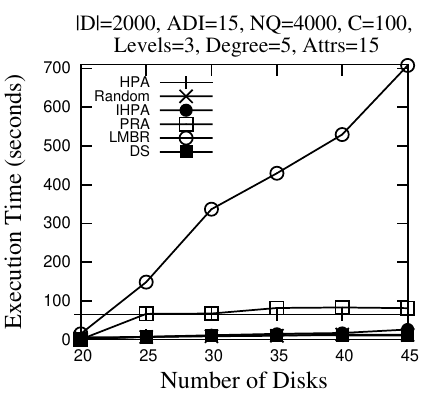}
\label{fig:snow3}
}
\vspace{-10pt}
\caption[]{Results of the Experiments on the Snowflake Dataset}
\label{fig:experiments2}
\vspace{-6pt}
\end{figure}

Figures~\ref{fig:snow1} and \ref{fig:snow3} show a set of experimental results for the Snowflake dataset. Each
of the plotted numbers corresponds to an average over 10 random query workloads. The data item
graph itself was generated with the following parameters: the number of levels in the graph 
was 3, the degree of each relation (the maximum number of tables it may join with) is set to 5,
and the number of attributes per table is set to 15. The total number of data items was 2000,
requiring a minimum of 20 partitions to store them. Note that we assume homogeneous data items in this
case. We plot the average query spans, and the average execution times
as the number of partitions increases from 20 to 45. 

\begin{figure}[t]
\centering
\subfigure[]{
\hspace{-.3in}
\includegraphics[scale=0.98]{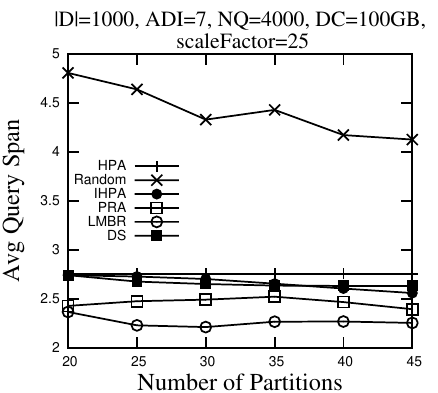}
\label{fig:het1}
}
\hspace{-.12in}
%\hspace{.1in}
%\subfigure[]{
%\includegraphics[scale=1.10, trim=0 0 10 0]{het2.pdf}
%\label{fig:het2}
%}
%\hspace{.1in}
\subfigure[]{
\includegraphics[scale=0.98]{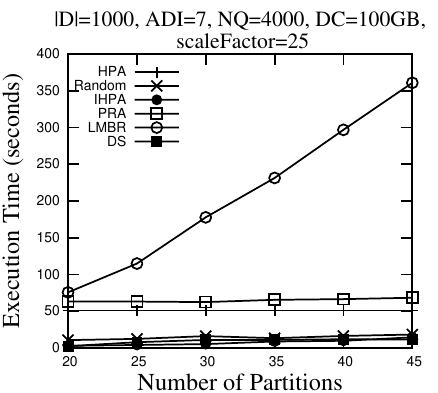}
\label{fig:het3}
}
\caption[]{Results of the Experiments on a TPC-H style Benchmark with unequal data item sizes. The relation sizes were calculated assuming a scale factor of 25.}
\label{fig:hetero}
\vspace{-8pt}
\end{figure}

We also conducted a similar set of experiments with 
heterogeneous data item sizes, where we generated TPC-H style queries with data item sizes 
adhering to the TPC-H benchmark. We chose the scale factor of 25, which means the highest data item 
size is 28GB and smallest data item size is 25KB. This results in a high skew among the table column 
sizes. Data item size is calculated as $Size(columnDatatype)*noRows$. The partition capacity was fixed at 100GB, and
we once again plot the average query spans and the average execution times
as the number of partitions increases from 20 to 45. 
The results are shown in Figures~\ref{fig:het1} and \ref{fig:het3}.

Our results here corroborate the results on the Random dataset. We once again see that LMBR 
performs the best, finding significantly better data layouts than the other algorithms. 
The performance differences are quite drastic with homogeneous data item sizes -- with 45 partitions, LMBR is able
to achieve an average query span of just 1.5, whereas the baseline HPA results in an average
span of 3.5. However, we observe that with heterogeneous data item sizes, the advantages of using
smart data placement algorithms are lower. With an extreme skew among the data item sizes,
the replication and data placement choices are very limited.

\subsubsection{ISPD98 Benchmark Dataset}
\begin{figure}[htb]
\centering
\hspace{-.1in}
\includegraphics[scale=1, trim=0 0 10 0]{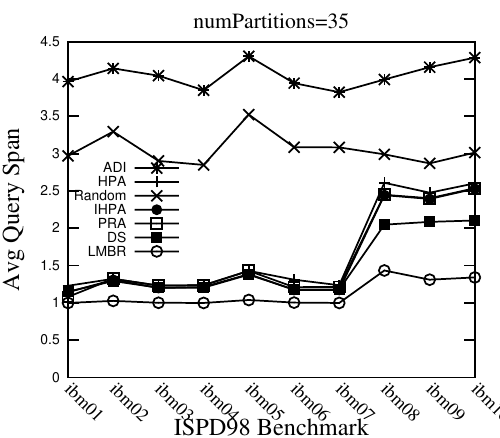}
\vspace{-10pt}
\caption[]{Results of the experiments on the first 10 hypergraphs, {\em ibm01, \ldots, ibm10}, from the ISPD98 Benchmark Dataset}
\label{fig:bench}
\vspace{-8pt}
\end{figure}

Finally, Figure \ref{fig:bench} shows the comparative results for first ten of hypergraphs from the 
ISPD98 Benchmark Suite, 
commonly used in the hypergraph partitioning literature. The number of hyperedges in the datasets range from 14111 to 75196 and number of nodes range from 12752 to 69429. Here we set the partition capacity so that exactly 20 partitions are sufficient to store the data items, and we plot the results with
number of partitions set to 35. The hypergraphs in this dataset tend to have fairly low densities, resulting
in low query spans. In fact, LMBR is able to achieve an average query span of close to the minimum
possible (i.e., 1) with 35 partitions. Most of the other algorithms perform about 20 to 40\% worse
compared to LMBR. 

These additional experiments further corroborate our claim that intelligent data placement with 
replication can significantly reduce the coordination overheads in data centers, and further that our
LMBR algorithm outperforms rest of the algorithms significantly.

\eat{
Since it is practically not possible to setup large number of partitions and measure each partition power
consumption precisely, using an accurate partition power model is more appropriate and feasible. However
it is possible to measure the single partition energy using multimeters and tools like Kill-A-Watt,
%\cite{killawatt}, 
but logging the measurements would need software tools to interface with voltmeter
or a multimeter. It is very much preferable to remove this dependency on hardware equipment, so we
studied various works \cite{mhd,DBLP:conf/mascots/HylickSRJ08,Greenawalt94modelingpower}
on partition energy profiling and power modelling. We draw important insights from these work and we
develop our own trace driven simulation framework with partition power modelling. Hylick et al.
\cite{hylick}, develop a simple methodology for creating accurate hard drive runtime energy models
through the use of easily obtainable data derived from published specifications and performance
measurements. Their work is important because of simplicity and accuracy of their partition power
estimation model. One can easily estimate the power components of the partition drive using the published
partition power and performance specifications. We directly use their model to estimate partition drive power
consumption estimation in our trace driven simulation framework.
}

%\section{Discussion}
%\label{sec:discussion}

\section{Conclusions}
In this paper, we solve the combined problem of data placement and replication, given a query
workload, to minimize the total resource consumption and by proxy, the total energy consumption, in
very large distributed or multi-site read-only data stores. Directly optimizing for either of these
metrics is likely infeasible in most practical scenarios because of the large number of factors
involved. We instead identify query span, the number of machines involved in executing a query, as
having a direct and significant impact on the total resource consumption, and focus on minimizing
the average query span for a given query workload. We formulated and analyzed the problems of data
placement and replica selection for this metric, and drew connections to several well-studied graph
theoretic concepts. We used these connections to develop a series of algorithms to solve this
problem, and our extensive experimental evaluation over several datasets demonstrated that our
algorithms can result in drastic reductions in average query spans. We are planning to extend our work in several different directions. As
we discussed earlier, we believe that temporal scheduling algorithms can be used to correct the load
imbalance that may result from optimizing for query span alone; although analysis tasks are
usually not latency sensitive, there are still often deadlines that need to be satisfied. We plan to
study how to incorporate such deadlines into our framework. We are also planning to study how to
efficiently track changes in the query workload nature online, and how to adapt the replication
decisions online. 
%We aim to minimize total work as
%opposed to response times, because complex and sophisticated analysis queries are response time
%in-sensitive. Our techniques can also result in reduced energy consumption and increased parallelism
%and concurrency of the overall system. We model the query workload as hypergraph and perform
%min-cuts to minimize the {\em query span}. We present series of replication algorithms to further
%minimize the min-cuts. These algorithms trade execution times with quality of the min-cuts. We find
%that in-graph replication techniques are more scalable than partiton-analyze-replicate based
%algorithms like LMBR, however LMBR gives the best reduction in {\em query span} compared to other
%techniques. One natural question that arises while we minimize the {\em query span} is, how do we
%address the issue of load balancing? Modelling the query workload as hypergraph gives us the
%flexibility of assigning weights to the vertices. We can either assign based on storage or load
%constraints of the partitions. Another flexibility that we have is in-terms of considering the
%temporal aspect of the queries where we can postpone the execution of certain queries for time being
%to guarantee load balancing. 

{
\small
\bibliographystyle{abbrv} 
\bibliography{writeup-cleaned} 

\begin{thebibliography}{10}

\bibitem{fullpaper}
\url{http://www.cs.umd.edu/~ashwin/fullpaper.pdf}.

\bibitem{hMETIS}
{hMETIS}: A hypergraph partitioning package,
  http://glaros.dtc.umn.edu/gkhome/metis/hmetis/overview.

\bibitem{MLPart}
{MLPart}, http://vlsicad.ucsd.edu/gsrc/bookshelf/slots/partitioning/mlpart/.

\bibitem{DBLP:conf/stoc/AlonST90}
N.~Alon, P.~D. Seymour, and R.~Thomas.
\newblock A separator theorem for graphs with an excluded minor and its
  applications.
\newblock In {\em STOC}, 1990.

\bibitem{274546}
C.~J. Alpert.
\newblock The {ISPD98} circuit benchmark suite.
\newblock In {\em Proc. of Intl. Symposium on Physical Design}, 1998.

\bibitem{socc2010}
H.~Amur, J.~Cipar, V.~Gupta, G.~Ganger, M.~Kozuch, and K.~Schwan.
\newblock Robust and flexible power-proportional storage.
\newblock In {\em SoCC}, 2010.

\bibitem{asahiro:swat96}
Y.~Asahiro, K.~Iwama, H.~Tamaki, and T.~Tokuyama.
\newblock Greedily finding a dense subgraph.
\newblock In {\em SWAT}, 1996.

\bibitem{boppana}
R.~B. Boppana.
\newblock Eigenvalues and graph bisection: An average-case analysis.
\newblock In {\em FOCS}, 1987.

\bibitem{384247}
A.~E. Caldwell, A.~B. Kahng, and I.~L. Markov.
\newblock Design and implementation of move-based heuristics for {VLSI}
  hypergraph partitioning.
\newblock {\em J. Exp. Algorithmics}, 5:5, 2000.

\bibitem{DBLP:conf/sigcomm/ChowdhuryZMJS11}
M.~Chowdhury, M.~Zaharia, J.~Ma, M.~I. Jordan, and I.~Stoica.
\newblock Managing data transfers in computer clusters with {Orchestra}.
\newblock In {\em SIGCOMM}, pages 98--109, 2011.

\bibitem{762819}
D.~Colarelli and D.~Grunwald.
\newblock Massive arrays of idle disks for storage archives.
\newblock In {\em Supercomputing}, 2002.

\bibitem{DBLP:journals/pvldb/CurinoZJM10}
C.~Curino, Y.~Zhang, E.~P.~C. Jones, and S.~Madden.
\newblock Schism: a workload-driven approach to database replication and
  partitioning.
\newblock {\em PVLDB}, 3(1):48--57, 2010.

\bibitem{Dittrich:2010:HMY:1920841.1920908}
J.~Dittrich, J.-A. Quian\'{e}-Ruiz, A.~Jindal, Y.~Kargin, V.~Setty, and
  J.~Schad.
\newblock Hadoop++: making a yellow elephant run like a cheetah (without it
  even noticing).
\newblock {\em PVLDB}, 3:515--529, September 2010.

\bibitem{Du20111224}
Z.~Du, J.~Hu, Y.~Chen, Z.~Cheng, and X.~Wang.
\newblock Optimized qos-aware replica placement heuristics and applications in
  astronomy data grid.
\newblock {\em Journal of Systems and Software}, 84(7):1224 -- 1232, 2011.

\bibitem{Economou06full-systempower}
D.~Economou, S.~Rivoire, and C.~Kozyrakis.
\newblock Full-system power analysis and modeling for server environments.
\newblock In {\em In Workshop on Modeling Benchmarking and Simulation (MOBS)},
  2006.

\bibitem{DBLP:journals/pvldb/EltabakhTOGKM11}
M.~Y. Eltabakh, Y.~Tian, F.~{\"O}zcan, R.~Gemulla, A.~Krettek, and
  J.~McPherson.
\newblock Cohadoop: Flexible data placement and its exploitation in hadoop.
\newblock {\em PVLDB}, 4(9):575--585, 2011.

\bibitem{Feige99thedense}
U.~Feige, G.~Kortsarz, and D.~Peleg.
\newblock The dense k-subgraph problem.
\newblock {\em Algorithmica}, 1999.

\bibitem{1055577}
H.~Ferhatosmanoglu, A.~S. Tosun, and A.~Ramachandran.
\newblock Replicated declustering of spatial data.
\newblock In {\em PODS}, 2004.

\bibitem{garey-johnson:79}
M.~Garey and D.~Johnson.
\newblock {\em {``Computers and Intractability: A Guide to the Theory of
  NP-Completeness''}}.
\newblock 1979.

\bibitem{1385494}
G.~Graefe.
\newblock Database servers tailored to improve energy efficiency.
\newblock In {\em Proceedings of EDBT workshop on Software engineering for
  tailor-made data management}, 2008.

\bibitem{DBLP:journals/corr/abs-0909-1784}
S.~Harizopoulos, M.~A. Shah, J.~Meza, and P.~Ranganathan.
\newblock Energy efficiency: The new holy grail of data management systems
  research.
\newblock In {\em CIDR}, 2009.

\bibitem{10.1109/CLOUD.2011.17}
L.-Y. Ho, J.-J. Wu, and P.~Liu.
\newblock Optimal algorithms for cross-rack communication optimization in
  mapreduce framework.
\newblock In {\em IEEE International Conference on Cloud Computing}, 2011.

\bibitem{Karypis99multilevelhypergraph}
G.~Karypis, R.~Aggarwal, V.~Kumar, and S.~Shekhar.
\newblock Multilevel hypergraph partitioning: Application in {VLSI} domain.
\newblock In {\em IEEE VLSI}, pages 69--529, 1999.

\bibitem{Karypis98multilevelk-way}
G.~Karypis and V.~Kumar.
\newblock Multilevel k-way hypergraph partitioning.
\newblock In {\em Proc. of DAC}, pages 343--348, 1998.

\bibitem{koyuturk05}
M.~Koyut\"urk and C.~Aykanat.
\newblock Iterative-improvement-based declustering heuristics for multi-disk
  databases.
\newblock {\em Information Systems}, 2005.

\bibitem{DBLP:conf/cidr/LangP09}
W.~Lang and J.~M. Patel.
\newblock Towards eco-friendly database management systems.
\newblock In {\em CIDR}, 2009.

\bibitem{DBLP:journals/pvldb/LangP10}
W.~Lang and J.~M. Patel.
\newblock Energy management for mapreduce clusters.
\newblock {\em PVLDB}, 3(1):129--139, 2010.

\bibitem{1740405}
J.~Leverich and C.~Kozyrakis.
\newblock On the energy (in)efficiency of hadoop clusters.
\newblock {\em HotPower}, 2009.

\bibitem{DBLP:journals/is/LiuS96}
D.-R. Liu and S.~Shekhar.
\newblock Partitioning similarity graphs: A framework for declustering
  problems.
\newblock {\em Information Systems}, 1996.

\bibitem{1577032}
H.~Meyerhenke, B.~Monien, and T.~Sauerwald.
\newblock A new diffusion-based multilevel algorithm for computing graph
  partitions.
\newblock {\em J. Parallel Distrib. Comput.}, 69(9), 2009.

\bibitem{sigmod2012jayanta}
J.~Mondal and A.~Deshpande.
\newblock Managing large dynamic graphs efficiently.
\newblock In {\em SIGMOD}, 2012.

\bibitem{journals/endm/NevesDOAU10}
T.~A. Neves, L.~M. de~A.~Drummond, L.~S. Ochi, C.~Albuquerque, and E.~Uchoa.
\newblock Solving replica placement and request distribution in content
  distribution networks.
\newblock {\em Electronic Notes in Discrete Mathematics}, 36:89--96, 2010.

\bibitem{1616815}
K.~Y. Oktay, A.~Turk, and C.~Aykanat.
\newblock Selective replicated declustering for arbitrary queries.
\newblock In {\em Euro-Par}, 2009.

\bibitem{Olston:2008:PLN:1376616.1376726}
C.~Olston, B.~Reed, U.~Srivastava, R.~Kumar, and A.~Tomkins.
\newblock Pig latin: a not-so-foreign language for data processing.
\newblock In {\em SIGMOD}, 2008.

\bibitem{Pavlo:2009:CAL:1559845.1559865}
A.~Pavlo, E.~Paulson, A.~Rasin, D.~J. Abadi, D.~J. DeWitt, S.~Madden, and
  M.~Stonebraker.
\newblock A comparison of approaches to large-scale data analysis.
\newblock In {\em SIGMOD}, 2009.

\bibitem{1006220}
E.~Pinheiro and R.~Bianchini.
\newblock Energy conservation techniques for disk array-based servers.
\newblock In {\em Supercomputing}, 2004.

\bibitem{abdulsubmission}
A.~Quamar, K.~A. Kumar, and A.~Deshpande.
\newblock Sword: Scalable workload-aware data placement for transactional
  workloads.
\newblock In {\em In submission}.

\bibitem{271619}
H.~D. Simon and S.-H. Teng.
\newblock How good is recursive bisection?
\newblock {\em SIAM J. Sci. Comput.}, 18(5):1436--1445, 1997.

\bibitem{grid2-ch19}
D.~Thain and M.~Livny.
\newblock Building reliable clients and servers.
\newblock In I.~Foster and C.~Kesselman, editors, {\em The Grid: Blueprint for
  a New Computing Infrastructure}. Morgan Kaufmann, 2003.

\bibitem{Thusoo}
A.~Thusoo, J.~S. Sarma, N.~Jain, Z.~Shao, P.~Chakka, S.~Anthony, H.~Liu,
  P.~Wyckoff, and R.~Murthy.
\newblock Hive: a warehousing solution over a map-reduce framework.
\newblock {\em PVLDB}, 2:1626--1629, August 2009.

\bibitem{Tosun97optimalparallel}
A.~A. Tosun and H.~Ferhatosmanoglu.
\newblock Optimal parallel {I/O} using replication.
\newblock In {\em ICPP}, 1997.

\bibitem{968054}
A.~S. Tosun.
\newblock Replicated declustering for arbitrary queries.
\newblock In {\em ACM symposium on Applied computing}, 2004.

\bibitem{1807194}
D.~Tsirogiannis, S.~Harizopoulos, and M.~A. Shah.
\newblock Analyzing the energy efficiency of a database server.
\newblock In {\em SIGMOD}, 2010.

\bibitem{citeulike:4882841}
T.~White.
\newblock {\em {Hadoop: The Definitive Guide}}.
\newblock {O'Reilly Media}, 1st edition, {June} 2009.

\bibitem{Wolfson97anadaptive}
O.~Wolfson, S.~Jajodia, and Y.~Huang.
\newblock An adaptive data replication algorithm.
\newblock {\em ACM TODS}, 22:255--314, 1997.

\bibitem{Wolfson:1991:MPR:103140.103146}
O.~Wolfson and A.~Milo.
\newblock The multicast policy and its relationship to replicated data
  placement.
\newblock {\em ACM TODS}, 16:181--205, March 1991.

\end{thebibliography}
}

\end{document}